\documentclass[aps,prb,twocolumn,superscriptaddress]{revtex4-1}

\usepackage{amsfonts}
\usepackage{amsmath}
\usepackage{amssymb}
\usepackage{graphicx}
\usepackage{color}
\usepackage{amsmath}
\usepackage{mathtools}
\usepackage{hyperref}
\usepackage{float}
\usepackage{cancel}
\usepackage{upgreek}

\newcommand{\ud}{\mathrm{d}}

\hypersetup{colorlinks,breaklinks,
            linkcolor=blue,urlcolor=blue,anchorcolor=blue,citecolor=blue}

\begin{document}

\title{Disconnection-Mediated Migration of Interfaces in Microstructures: \\
II. diffuse interface simulations}

\author{Marco Salvalaglio}
\affiliation{Institute of Scientific Computing, TU Dresden, 01062 Dresden, Germany}
\affiliation{Dresden Center for Computational Materials Science, TU Dresden, 01062 Dresden, Germany}
\affiliation{Hong Kong Institute for Advanced Study, City University of Hong Kong, Hong Kong SAR, China}
\author{David J. Srolovitz}
\affiliation{Department of Mechanical Engineering, The University of Hong Kong, Pokfulam Road, Hong Kong SAR, China}
\author{Jian Han}
\affiliation{Department of Materials Science and Engineering, City University of Hong Kong, Hong Kong SAR, China}


\begin{abstract}
The motion of interfaces is an essential feature of microstructure evolution in crystalline materials. While atomic-scale descriptions provide mechanistic clarity, continuum descriptions are important for understanding microstructural evolution and upon which microscopic features it depends. We develop a microstructure evolution simulation approach that is linked to the underlying microscopic mechanisms of interface migration. We extend the continuum approach describing the disconnection-mediated motion of interfaces introduced in Part I [Han, Srolovitz and Salvalaglio, 2021] to a diffuse interface, phase-field model suitable for large-scale microstructure evolution. A broad range of numerical simulations showcases the capability of the method and the influence of microscopic interface migration mechanisms on microstructure evolution. These include, in particular, the effects of stress and its coupling to interface migration which arises from disconnections, showing how this leads to important differences from classical microstructure evolution represented by mean curvature flow.
\end{abstract}

\maketitle

\section{Introduction}

Interfaces are arguably the most important elements of the microstructure of materials.
Since they play a key role in material properties, their control is essential to the design of engineering materials.
Simulation of microstructure evolution is complex because of the wide range of length- and time-scales involved\cite{Rollett2015}.
For example, while the structure and dynamics of most grain boundaries (GBs) can only be elucidated by a combination of atomistic and crystallographic approaches, elastic interactions amongst grains/GB in a polycrystalline microstructure and microstructure evolution must be based upon coarse-grained or macroscopic approaches appropriate for large length- and time-scales.  

Ideally, approaches that provide comprehensive descriptions of interfaces in crystalline systems should retain details of different length- and time-scales. 
At the same time, they should be versatile to cope with the complexity of experimental systems. 
For example, they should be able to describe a wide range of complex interface geometries, allowing for parameterisations appropriate for large length- and time-scales and coupling of different physical effects occurring simultaneously.

Approaches based upon the underlying interface dynamics have greater potential for comprehensive descriptions of microstructure evolution than those built from either purely atomistic or continuum approaches.
Such approaches, for example, may be built upon the main carriers of interface dynamics in crystalline materials; i.e., \textit{disconnections} \cite{bollmann1970general,hirth1973grain,balluffi1982csl,hirth2006disconnections,hirth2007spacing,Han2018}  which are line defects with dislocation and step characters. 
This description may be abstracted in the form of a general, disconnection-mediated,  interface equation of motion (EOM)\cite{SUTTONBOOK,Zhang2017,Zhang2018,Zhang2021,Han2021}. 
It extends classical, purely continuum models for interface motion, such as mean-curvature flow or the motion by the Laplacian of the interface curvature \cite{Mullins1957,Mullins1959,Doherty1997}, that are widely applied in materials science for microstructure evolution. 

In Part I of this paper\cite{Han2021}, we introduced a disconnection-based EOM approach for arbitrarily curved interfaces, overcoming severe limitations of previous formulations for application to general microstructure evolution.
In this paper, we propose a general framework for microstructure evolution, encoding this description in a continuum framework that easily and simultaneously handles nontrivial interface geometries/morphologies, topological changes, anisotropy in kinetic and thermodynamic properties, and multiple interfaces. 
Our approach is based upon the well-established diffuse interface, phase-field (PF) model \cite{chen2002phase,boettinger2002phase,Steinbach_2009,Li2009,provatas2011phase}. 
We note that several PF models have previously been proposed to study grain boundaries, coherent interfaces, and microstructure evolution. 
These focus either on multi-order-parameter approaches\cite{Chen1994,steinbach1996phase,Moelans2008,Steinbach_2009,DARVISHIKAMACHALI20122719,Toth2015,DIMOKRATI2020147} or the incorporation of additional fields that account for local crystal orientations\cite{KOBAYASHI2000141,WARREN20036035,Herve2012,Korbuly2017}. 
Both approaches for macroscopic modelling of microstructure have advantages for specific, targeted applications. 
Here, we consider single and multi phase-field models to track the motion of interfaces based upon the underlying disconnection mechanisms~\cite{Zhang2017,Zhang2018,Han2018,Han2021}. 
To achieve this goal, we exploit phase-field models originally designed for (interface) mean curvature flow and their extension to systems with many interfaces~\cite{Rubinstein1992,Li2009,Brassel2011,LEE201635,bretin2017new,BRETIN2018324}. 
We then extend these to include disconnection dynamics, focusing on a wide range of driving forces that act directly on disconnections with the aim of reproducing the sharp-interface EOM for arbitrarily curved interface \cite{Han2021}, such as the effects of externally applied and self-stress, chemical potential jumps across the interface, and capillarity, providing a convenient and versatile framework for describing a mechanistically-appropriate and general model for interface dynamics. As a central result, we demonstrate interface evolution effects which arise from disconnections, such as anisotropic shapes also in the presence of isotropic interface energy and mobility, growth of grains due to an external applied stress, grain translation and topological changes.

This paper is organised as follows. 
In Sect.~\ref{sec:cmodel}, we briefly review the basics of the continuum kinetic theory leading to an equation of motion for crystal interfaces based on disconnections mechanics; more details are in the preceding companion paper \cite{Han2021}. 
In Sect.~\ref{sec:PF}, we develop a diffuse interface approach that incorporates disconnection dynamics-based interface migration, while its extension to an arbitrary number of different interfaces is reported in~\ref{sec:multiPF}. 
In Sect.~\ref{sec:num}, we illustrate this approach through a varied set of numerical simulations exploiting simple integration schemes.

\section{Continuum Description}
\label{sec:cmodel}

\noindent

\begin{figure}
   \includegraphics[width=1\columnwidth]{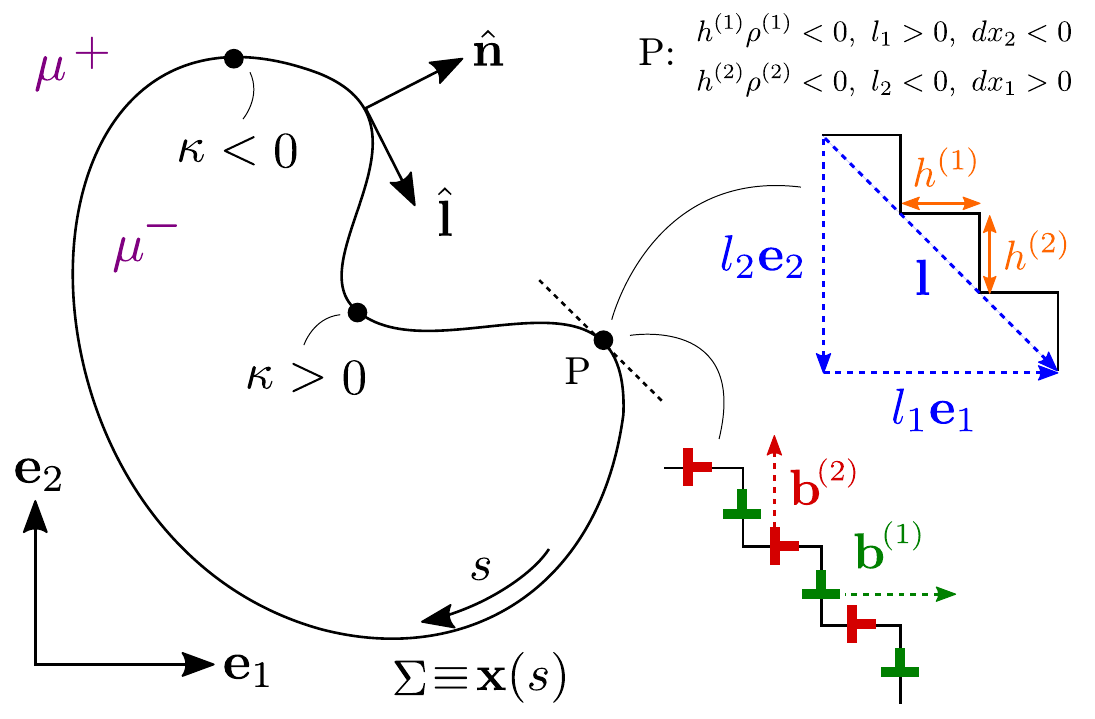} 
   \caption{Illustration of the variables and parameters in the continuum model (see Sect.~\ref{sec:cmodel}).
}
    \label{fig:figure1}
\end{figure}

We consider a continuum representation of interfaces accounting for steps and dislocations as disconnections (see  Part I\cite{Han2021}). 
For  simplicity of presentation, we focus on the minimal system encoding two interface references (each with a single type of disconnection; for extensions see \onlinecite{Han2021}).  
Figure~\ref{fig:figure1} illustrates the objects entering the continuum model. 
In brief, we consider a curve $\Sigma$ in the $\mathbf{e}_1$-$\mathbf{e}_2$ plane 
\begin{equation}\label{eq:el}
\mathbf{x}(s) = \left(
\begin{array}{c}
x_1(s) \\ x_2(s)
\end{array}
\right), 
\end{equation}
parametrised by $s$, with 
\begin{equation}
\begin{split}
\mathbf{l}(s)
=& \frac{\ud \mathbf{x}}{\ud s}
= \left(\begin{array}{c}
\ud x_1/\ud s \\ \ud x_2/\ud s
\end{array}\right)
=
\left(
\begin{array}{c}
l_1 \\ l_2
\end{array}
\right), \\
\hat{\mathbf{n}}(s)
=&\frac{1}{|\mathbf{l}|}
\left(
\begin{array}{c}
-l_2 \\ l_1
\end{array}
\right),
\end{split}
\end{equation}
its tangent vector and normal vector, respectively (the hat  denotes a normalized vector). 
The disconnection lines lie along $\mathbf{e}_3=\mathbf{e}_1 \times \mathbf{e}_2$.
Equation~\eqref{eq:el} can be rewritten in terms of the step heights $h^{(k)}$ and disconnection densities $\rho^{(k)}(s)dL$ (moving along the curve in direction $\mathbf{e}_k$, with $k=1,2$ and $dL$ the arc along $\Sigma$) through
\begin{equation}
\hat{\mathbf{l}}=\left(
\begin{array}{c}
-h^{(2)}\rho^{(2)} \\ h^{(1)}\rho^{(1)}
\end{array}
\right),
\end{equation}
adopting the convention $h^{(k)}>0$. 
$\rho^{(1)} \gtreqless 0$ corresponds to $dx_2 \gtreqless 0$ and $\rho^{(2)} \gtreqless 0$ corresponds to $dx_1 \lesseqgtr 0$ (Fig.~\ref{fig:figure1} shows an example of this at Point P on $\Sigma$). 

The dislocation character or Burgers vector is  $\mathbf{b}^{(k)}=b_k\mathbf{e}_k$ and the Burgers vector density along 
 a unit arc length is
\begin{equation}
\begin{split}
\rho^{(m)}b^{(m)}=&\frac{b^{(m)}}{h^{(m)}}h^{(m)}\rho^{(m)}
=(-1)^n\beta^{(m)}\hat{l}_n,
\end{split}
\end{equation}
where $(m,n)=(1,2)$ or $(2,1)$, and $\beta^{(m)}$ is the shear-coupling factor for the interface with tangent vector $\mathbf{e}_m$.

The interface evolution law is based on the disconnection density along the curve and the local driving forces \cite{Han2021}
\begin{equation}
\mathbf{f}=\mathbf{f}_{\rm s}+\mathbf{f}_{\rm d}+\psi \hat{\mathbf{n}}=(\Gamma \kappa+ \tau\Lambda+ \psi )\hat{\mathbf{n}}.
\label{eq:totdrivingforce}
\end{equation}
$\mathbf{f}_{\rm s}$ is the gradient descent of the interface energy 
\begin{equation}
\mathbf{f}_{\rm s}=-\frac{\delta E[\mathbf{x}(s)]}{\delta \mathbf{x}(s)}=\Gamma \kappa \hat{\mathbf{n}},
\end{equation}
where
\begin{equation}\label{E}
E[\mathbf{x}(s)] 
= \int_\Sigma \gamma(s) |\mathbf{l}(s)| \,\ud s.
\end{equation} 
$\gamma(s)$ is the interface energy density (alternatively expressed as a function of the interface inclination angle $\phi$; $\gamma(s) = \gamma(\phi(s))$ in 2D), $\Gamma=\gamma+\gamma_{,\phi\phi}$ is the interface stiffness and $\kappa$ is the local curvature of the interface. 
$\mathbf{f}_{\rm d}$ is the Peach-Koehler (PK) force acting on the disconnection Burgers vector from the total stress $\boldsymbol{\upsigma}$ at that interface position, namely
$\mathbf{f}_\text{d} 
= (\boldsymbol{\upsigma}\mathbf{b}) \times \mathbf{e}_3.$
Assuming disconnection can only move conservatively (i.e., by glide), this reduces  to 
$\mathbf{f}_\text{d} = \tau \Lambda \hat{\mathbf{n}}$,
where $\tau \equiv \sigma_{12}$ the total shear stress and
\begin{equation}\label{defLambda}
\Lambda
\equiv \beta^{(2)} - \beta^{(1)}
= \frac{b^{(2)}}{h^{(2)}} - \frac{b^{(1)}}{h^{(1)}}.
\end{equation}
$\psi$ in Eq.~\eqref{eq:totdrivingforce} accounts for the  chemical potential ($\mu$) jump  across the interface $\psi=\mu^{+}-\mu^{-}$ (``$\pm$'' denotes the side of interface to/from which $\hat{\mathbf{n}}$ points). For a heterophase interface, $\psi$ is the difference in free energy per atom of the phases on the $\pm$ sides of the interface. 

Assuming that disconnection dynamics are overdamped, we write the evolution law for $\mathbf{x}(s)$ as
\begin{equation}
\dot{\mathbf{x}}(s) = \mathbf{M}\mathbf{f} = (\Gamma \kappa + \tau\Lambda + \psi) \mathbf{M}\hat{\mathbf{n}}(s),
\label{eq:modeleq}
\end{equation}
with 
\begin{equation}
\mathbf{M}
\equiv 
\left(\begin{array}{cc}
M^{(1)} & 0 \\ 
0 & M^{(2)}
\end{array}\right).
\label{eq:modelM}
\end{equation}
Curvatures, jumps in  chemical potential across  interfaces, external stresses, and disconnection stress-sources in Eq.~\eqref{eq:modeleq}, vary throughout the microstructure (which evolves with time). Limiting cases can be easily recovered; e.g. in the absence of  dislocation character or chemical potential jumps this reduces to anisotropic mean curvature flow,  the dominant roles of  stress for flat interfaces with non-zero dislocation character, or to heterophase interface motion with  $\psi \neq 0$ (in the absence of a stress). 
The balance between these terms  depends both on the physical situation as well as the parameters  $\gamma^{(k)}$, $M^{(k)}$, $\beta^{(k)}$.

We consider the specific interface mobility and energy density anisotropy associated to the orientation of the reference interfaces\cite{Han2021}
\begin{equation}
M(\phi)
= \hat{\mathbf{n}}\cdot\mathbf{M}\hat{\mathbf{n}}
= M^{(1)}\cos^2\phi + M^{(2)}\sin^2\phi,
\label{eq:mob}
\end{equation}
\begin{equation}
\gamma(\phi)
= \gamma^{(2)} |\cos\phi|
+ \gamma^{(1)} |\sin\phi|.
\label{eq:gammaaniso}
\end{equation}
This interface energy is cusped and corresponds to  preferred interface orientations;  i.e., to equilibrium interface facets  \cite{Wulff1901,Herring1951}. 
Such singularities in $\gamma(\phi)$ leads to sharp corners in the equilibrium faceted interface profiles;  this is the strong-anisotropy regime. This is a well-known condition that may pose issues for continuum approaches better suited for continuous profiles \cite{taylor1998diffuse,Spencer2004}. 
To avoid this issue, we here simply regularise the interface energy density as
\begin{equation}
\gamma_\eta(\phi)
= \gamma^{(2)} R_\eta(\cos\phi)
+ \gamma^{(1)} R_\eta(\sin\phi),
\label{eq:gammaaniso_R}
\end{equation}
where
\begin{equation}\label{regular}
R_\eta(z)=\frac{\eta}{5}\bigg[\ln(2)+\ln(1+\cosh(5z/\eta)) \bigg] \stackrel{\eta\rightarrow 0}{\approx} |z|,
\end{equation}
is a smooth approximation for $|z|$ that provides a localized corner smoothing\cite{Herty2007} and formally converges to the nominal Wulff shape for $\eta \rightarrow 0$. Several other regularisations have been proposed, including e.g. an additional energy term or similarly enforcing rounding at cusps of $|z|$ with some parametrization \cite{taylor1998diffuse,Debierre2003,Spencer2004,Wan2018,Philippe2021,Han2021}. 
Examples of shapes and dynamics obtained with different values of the regularisation parameter $\eta$ are shown below, where we illustrate the convergence to faceted shapes for small $\eta$.

The evolution of $\mathbf{x}(s)$ is  determined by integration of Eq.~\eqref{eq:modeleq} given the stress $\tau$. 
This stress can be  separated into  contributions from an  external shear stress $\tau_{\rm ext}$ and those generated by the disconnections themselves along the entire interface profile $\mathbf{x}(s)$, $\tau_{\rm self}(s)$. 
The latter, assuming an isotropic and homogeneous medium, can be computed as 
\begin{equation}\label{s12curve_s}
\tau_\text{self}(s)
= \beta^{(1)} I^{(1)}_\Sigma(s) + \beta^{(2)} I^{(2)}_\Sigma(s), 
\end{equation}
where 
\begin{align}\label{s12curve_s2}
I^{(m)}_\Sigma(s)
&= \frac{G}{2\pi(1-\nu)} \int_\Sigma 
\Bigg\{\left(\frac{\ud x_n}{\ud s}\right)_{s=s_0} 
\dfrac{x_m(s) - x_m(s_0)}{\varrho_a^2} 
\nonumber\\
&\times \left[
1 - \dfrac{2 \big(x_n(s) - x_n(s_0)\big)^2}{\varrho_a^2}
\right]\Bigg\} \ud s_0,
\end{align}
$(m, n) \in \{(1, 2),(2, 1)\}$, $\varrho_a^2 \equiv [x_1(s) - x_1(s_0)]^2 + [x_2(s) - x_2(s_0)]^2 + a^2$, $G$ and $\nu$ are the shear modulus and Poisson ratio, and $a$ encodes the disconnection core size\cite{cai2006non}. 

Note that with this approach we assume that the system is in elastic equilibrium. We exploit known elastic fields at (mechanical) equilibrium for disconnections/stress sources and evaluate their contribution at any point. Since we consider linear elasticity, the superposition principle holds for any stress sources (e.g., external/applied stresses). 
Stress sources associated with misfit may also be included by explicitly solving the mechanical equilibrium equations in the presence of eigenstrains. 
Such cases are not considered here.

The following reduced scales are adopted: $\tilde{\mathbf{x}} = \mathbf{x}/\alpha$ (same for other length quantities) and $\Delta\tilde{t} = \Delta t M_0\gamma_0/\alpha^2$, where $\alpha$ is the Displacement-Shift-Complete (DSC) lattice parameter commonly used in bicrystallography (as widely used to describe disconnection step heights and the Burgers vector norm~\cite{Han2021}).

\section{Diffuse Interface Modeling of a Single Interface}
\label{sec:PF}

\begin{figure}
   \includegraphics[width=1\columnwidth]{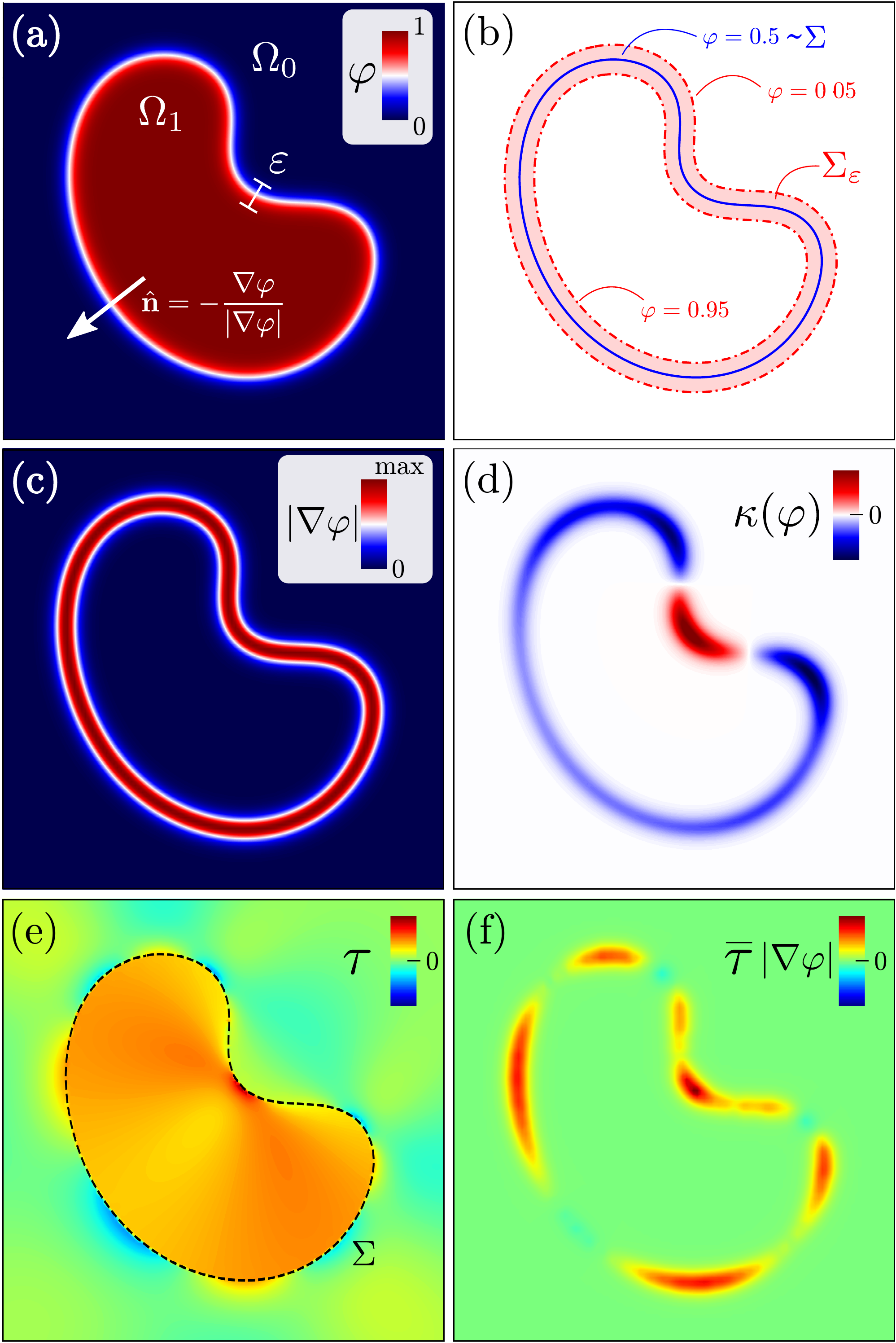} 
   \caption{Diffuse interface description of a sharp interface $\Sigma$ from Fig.~\ref{fig:figure1} in the phase-field model: (a) $\varphi(\mathbf{r})$, (b) selected $\varphi$ level sets ($\varphi=0.5$ corresponds to $\Sigma$), (c) $|\nabla \varphi|$, (d) a diffuse-interface representation of the local interface curvature of $\Sigma$ (see Eq.~\eqref{eq:ac_iso}), $\kappa(\varphi)$, (e) $\tau(\mathbf{r})$ (Eq.~\eqref{s12curve_s} evaluated at every point $\mathbf{r} \in \Omega$ instead of $s \in \Sigma$), for $\beta^{(1)}=1$, $\beta^{(2)}=0$ and $\tau_{\rm ext}=0$),  and (f) $\bar{\tau}(\mathbf{r})|\nabla \varphi|$ (Eq.~\eqref{eq:levelset} with $\tau(s)$ as in (e) on $\Sigma$). 
   The computational cell (reduced) size is $20$ and $\tilde{\varepsilon}=1$.}
    \label{fig:figure2}
\end{figure}

We now present a model for the evolution of arbitrary crystal interface shapes based on Eq.~\eqref{eq:modeleq} within a diffuse interface framework. 
We consider a phase-field model that easily accommodates mean curvature flow (shape evolution that minimises the interface area/energy). 
It tracks interface  ($\Sigma$, Fig.~\ref{fig:figure1}) evolution implicitly through an auxiliary order parameter which changes smoothly across the interface. 
This order parameter is a smooth function $\varphi(\mathbf{r})$ with $\mathbf{r} \in \Omega=\Omega_0 \cup \Omega_1 \cup \Sigma_\varepsilon$, that describes two phases, $\varphi=1$ for $\mathbf{r} \in \Omega_1$, $\varphi=0$ for $\mathbf{r} \in \Omega_0$, with a continuous transition in between (i.e., for $\mathbf{r} \in \Sigma_\varepsilon$) - see Fig.~\ref{fig:figure2}a-\ref{fig:figure2}c. 
$\varepsilon$ is a parameter that scales the diffuse interface width. 
$\varphi$ is determined from the minimization of a free energy functional that approximates the interface energy as introduced in Ref.
\onlinecite{Tor2009},
\begin{equation}
F[\varphi]=\int_\Omega \gamma(\hat{\mathbf{n}}) \left( \frac{\varepsilon}{2} |\nabla \varphi|^2  + \frac{1}{\varepsilon}H(\varphi) \right) \ud \mathbf{r},
\label{eq:energy}
\end{equation}
where $\gamma(\hat{\mathbf{n}})$ is an orientation-dependent interface energy density, $\hat{\mathbf{n}}=-\nabla \varphi / |\nabla \varphi| $ is the interface normal, and $H(\varphi)=18\varphi^2(1-\varphi)^2$ is a double well potential vanishing in the bulk phases. 
$|\nabla \varphi|^2$ makes the interface diffuse, while $H(\varphi)$ enforces the stability of the phases  (i.e., $\varphi=0,1$); 
their competition leads to a stable interface profile 
\begin{equation}
\varphi(\mathbf{r})=\frac{1}{2}\left[1-\tanh \left(\frac{3 d(\mathbf{r})}{\varepsilon}\right)\right],
\label{eq:tanh}
\end{equation}
where $d(\mathbf{r})$ is a signed distance from the 0.5 level set of $\varphi$; this contour approximates the corresponding sharp interface, $\Sigma$ in Fig.~\ref{fig:figure1}. With $\gamma(\hat{\mathbf{n}})$ multiplying both terms in \eqref{eq:energy} the interface thickness is independent of the interface orientation; this is a convenient feature for both  general numerical approaches (see detailed discussions in Ref. \onlinecite{Tor2009} and applications, e.g., in Refs.~\onlinecite{albani2019competition,Sal2015b,SalvalaglioDDCHaniso}).
Coefficients entering \eqref{eq:energy} and \eqref{eq:tanh} ensure that $F[\varphi] \approx E[\mathbf{x}(s)]$ for relatively small $\varepsilon$ (see, e.g., asymptotic analysis in Ref.~\onlinecite{Ratz2006}).

This framework conveniently describes mean curvature flow by computing $\dot{\varphi}$ as $L^2$-gradient flow of $F[\varphi]$; i.e., the Allen-Cahn equation \cite{AC1979,Li2009}. 
For isotropic interface energies $\gamma(\hat{\mathbf{n}})\equiv\gamma_0$ and mobilities $M(\hat{\mathbf{n}})\equiv M_0$, it yields
\begin{equation}
\dot{\varphi}=-\frac{M_0}{\varepsilon}\frac{\delta F}{\delta \varphi}=\frac{M_0\gamma_0}{\varepsilon}
\underbrace{\bigg[\varepsilon \nabla^2\varphi-\frac{1}{\varepsilon}H'(\varphi) \bigg]}_{\kappa(\varphi)},
\label{eq:ac_iso}
\end{equation}
with $\kappa(\varphi)$ a diffuse-interface representation of the interface curvature (see Fig.~\ref{fig:figure2}d).
Formally, Eq.~\eqref{eq:ac_iso}  asymptotically converges ($\varepsilon \rightarrow 0$) to isotropic mean curvature flow \cite{evans1992phase,Tor2009,Li2009}: $\dot{\mathbf{x}}(s)=M_0\gamma_0 \kappa \hat{\mathbf{n}}$. 

To this point, we have focused on the standard PF model to reproduce mean curvature flow. To account for new aspects related to driving forces associated with the total stress and chemical potentials defined for $\Sigma$, we add a term to the phase field evolution law for the advection of $\varphi$. 
Similar terms are common to describe translation of the interfaces by a prescribed velocity within phase field model, for instance when describing solidification and crystal growth \cite{medvedev2013simulating,rojas2015phase,qi2017modeling,albani2019competition}.
In practice, we consider the 0.5 level set of $\varphi$ as $\Sigma$ (see also Fig.~\ref{fig:figure2}). 
We then compute the additional velocity term $v_\Sigma=\mathbf{M}(\mathbf{f}_{\rm d}+\psi\hat{\mathbf{n}})$ on $\Sigma$ and extend it within the phase-field interface, i.e. in $\Sigma_\varepsilon$, obtaining a velocity $v(\hat{\mathbf{n}})$ constant along $\hat{\mathbf{n}}$ such that $\dot{\varphi}=\Phi=v(\hat{\mathbf{n}})|\nabla\varphi|$ approximates the motion of $\Sigma$ dictated by $v_\Sigma$; this occurs as $v(\hat{\mathbf{n}}) |\nabla \varphi| \approx v\delta_\Sigma$ for $\varepsilon \rightarrow 0$, where the delta function $\delta_\Sigma$ identifies the surface, as commonly exploited in level-set and diffuse domain approaches \cite{sethian1999level,osher2006level,li2009solving}. 
Here, the external stress $\tau_{\rm ext}$ and chemical potential jump $\psi$ are constants such that the conditions for advecting  $\varphi$ are met. 
On the other hand, $\tau_{\rm self}(s)$ is defined only on $\Sigma$ such that we must first compute the line integrals in Eq.~\eqref{s12curve_s} on the $\varphi \sim 0.5$ contour (see Eq.~\eqref{s12curve_s2} and Fig.~\ref{fig:figure2}e). 
The extension of the resulting $\tau_{\rm self}(s)$ within $\Sigma_\varepsilon$ (constant along $\hat{\mathbf{n}}$), can be achieved as the stationary solution of  \cite{sethian1999level,osher2006level}
\begin{equation}\label{eq:levelset}
\begin{split}
    \partial_{p}\bar{\tau}_{\rm self}&=S(\varphi-0.5) \hat{\mathbf{n}} \cdot \nabla \bar{\tau}_{\rm self}, \\
    S(z)&=\frac{z}{\sqrt{z^2+\delta^2}},\\
     \bar{\tau}_{\rm self}(\mathbf{r})&=\tau_{\rm self}(\mathbf{x}) \ \ \text{on}\ \ \Sigma\ : \{\mathbf{r}=\mathbf{x}\},
\end{split}
\end{equation}
where $p$ is a pseudo-time and $S(z)$ is a regularised sign function with small parameter $\delta=10^{-6}$, avoiding numerical divergences far away from the interface.   
Equation~\eqref{eq:levelset}  extends $\bar{\tau}_{\rm self}$ along the interface (positive and negative) normal. 
$\bar{\tau}_{\rm self}|\nabla\varphi|$ is  illustrated in Fig.~\ref{fig:figure2}f for $\beta^{(1)}=1$, $\beta^{(2)}=0$. 
Note that in this approach, we do not need to solve for the elastic fields concurrent  with the phase fields. 
We recall that through Eq.~\eqref{s12curve_s}, \eqref{s12curve_s2} and \eqref{eq:levelset}, the elastic fields of the dislocations and the external stress are included assuming   mechanical equilibrium.

The complete diffuse interface expression of Eq.~\eqref{eq:modeleq},  including anisotropy ($\gamma(\hat{\mathbf{n}})$, $M(\hat{\mathbf{n}})$) and  advection, is 
\begin{equation}
\begin{split}
\dot{\varphi}=&-\frac{M (\hat{\mathbf{n}})}{\varepsilon}\bigg[ \bigg(\frac{\delta F}{\delta \varphi}\bigg)+\varepsilon|\nabla \varphi|\bigg((\bar{\tau}_{\rm self}+\tau_{\rm ext})\Lambda+\psi \bigg) \bigg] \\
%
%
%
\approx &\frac{M (\hat{\mathbf{n}})}{\varepsilon} \bigg[\varepsilon  \nabla \cdot \bigg( \gamma(\hat{\mathbf{n}}) \nabla \varphi + |\nabla \varphi|^2 \mathbf{P} \nabla_{\hat{\mathbf{n}}} \gamma(\hat{\mathbf{n}}) \bigg)\\
&-\frac{\gamma(\hat{\mathbf{n}})}{\varepsilon}H'(\varphi) + \varepsilon|\nabla \varphi|\bigg((\tau_{\rm self}+\tau_{\rm ext})\Lambda+\psi \bigg)\bigg], 
\label{eq:ac}
\end{split}
\end{equation}
with $[\mathbf{P}]_{ij}=\delta_{ij}-\hat{{n}}_i\hat{{n}}_j$ \cite{Tor2009}, $\nabla_{\hat{\mathbf{n}}}$ representing the gradient with respect to the components of the normal vector, and exploiting the asymptotic result $(1/\varepsilon)H(\varphi)\stackrel{\varepsilon \rightarrow 0 }{\approx} (\varepsilon/2)|\nabla \varphi|^2$. \cite{Tor2009,Sal2015b,SalvalaglioDDCH,SalvalaglioDDCHaniso}
Anisotropic quantities may be expressed as functions of $\phi$:  $M(\hat{\mathbf{n}}) \equiv M(\phi)$  (Eq.~\eqref{eq:mob}) and $\gamma(\hat{\mathbf{n}}) \equiv \gamma(\phi)$  (Eq.~\eqref{eq:gammaaniso_R}), with $\phi=\arctan(\hat{n}_2/\hat{n}_1)=\arctan((\nabla \varphi)_2/(\nabla \varphi)_1)$. 

The equations reported above exactly recover the targeted sharp-interface dynamics (Sect.~\ref{sec:cmodel}, Ref.~\onlinecite{Han2021}) in the limit $\varepsilon \rightarrow 0$. While this condition cannot be realized in simulations (finite $\varepsilon$ is required), convergence ensures that a numerical approximation of the sharp-interface limit within the selected error limit can be achieved with a relatively small $\varepsilon$. 
$\varepsilon$ is typically chosen to be at least one order of magnitude smaller than the domain size characterised by the extension of the interface/phases described by $\varphi$. 
Similarly to any investigation based on phase-field simulations, this is how we will perform simulation exploiting the model illustrated in this and in the following sections.

\section{Diffuse Interface Modelling of Many Interfaces}
\label{sec:multiPF}

\begin{figure}[t!]
   \includegraphics[width=1\columnwidth]{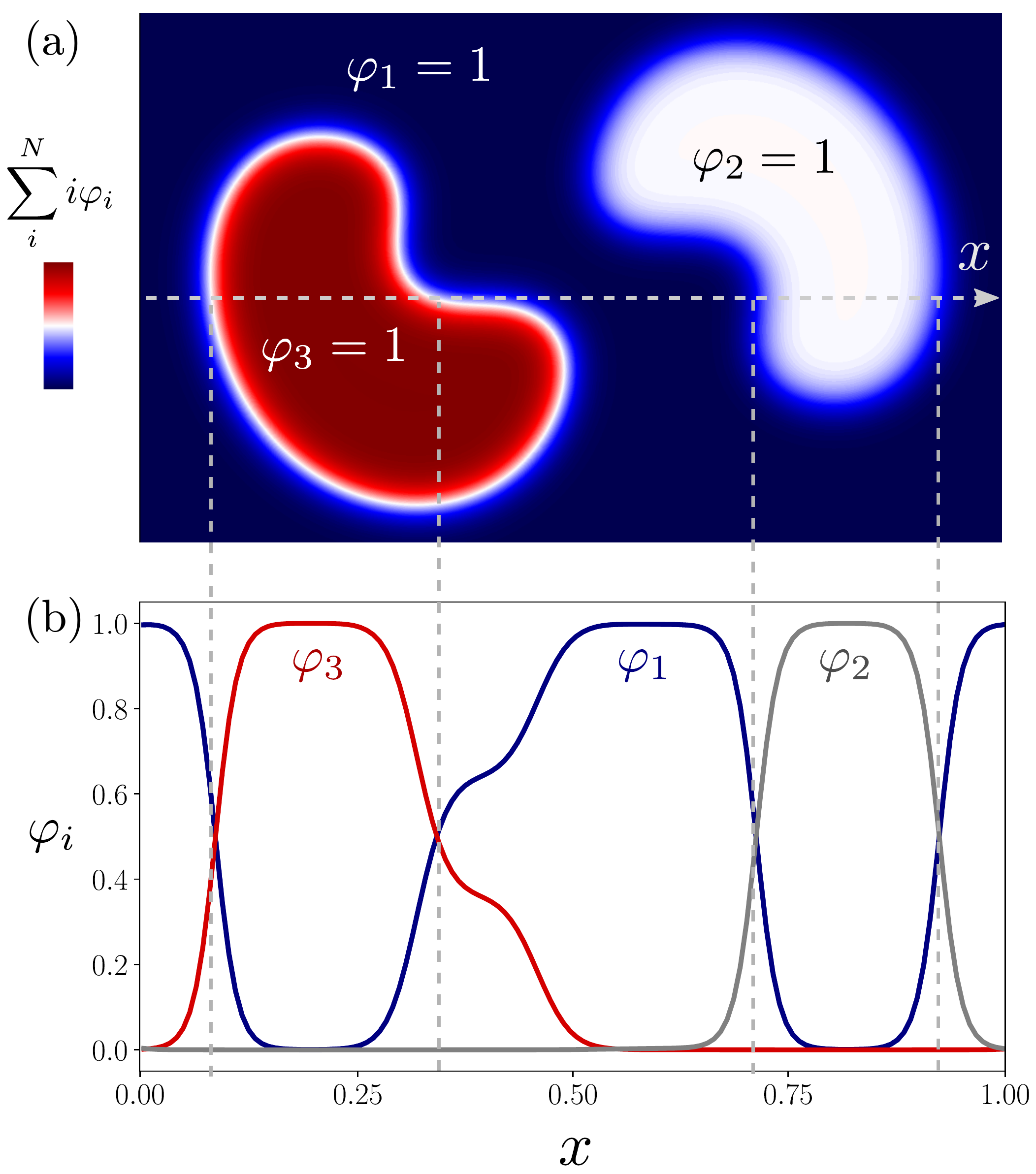} 
   \caption{Illustration of three phase fields in the multi phase-field model (Sect.~\ref{sec:multiPF}). (a) Color map illustrating the three phase fields. (b) Values of $\varphi_i$ along the dashed arrow ($x$) in  (a). Note that $\sum_i\varphi_i=1$ is  everywhere satisfied.}
    \label{fig:figure3}
\end{figure}

While the model discussed in Sect.~\ref{sec:PF} describes a single interface and directly translates the continuum description outlined in Sect.~\ref{sec:cmodel}, we now extend it to multi-phase systems with multiple interface types. 
For example, this description is necessary to describe grain boundaries between different grains in an anisotropic material (each grain orientation is described as a separate phase). 
Consider a set of phase fields $\varphi_i$ with $i=1,...,N$, each associated with an energy functional (as in Eq.~\eqref{eq:energy}) which satisfy $\sum_i^N \varphi_i=1$ (see Fig.~\ref{fig:figure3}). 
We write the total energy of the system as\cite{garcke1998anisotropic,garcke1999multiphase,Ratz2006,Tor2009,bretin2017new,BRETIN2018324}, 
\begin{equation}\label{eq:Fmulti}
    F_{\rm multi}[\{\varphi_i\}]=\frac{1}{2}\sum_i^N\int_\Omega \gamma_i\bigg(\frac{\varepsilon}{2}|\nabla \varphi_i|^2+\frac{1}{\varepsilon} H(\varphi_i)  \bigg) d\mathbf{r}.
\end{equation}
with $\gamma_i\equiv\gamma(\hat{\mathbf{n}}_i)$. The evolution of $\varphi_i$ is given by the $L^2$ gradient flow of $F_{\rm multi}$  with a Lagrange multiplier $\lambda$ that drives the gradient flow towards $\sum_i^N \varphi_i=1$; i.e., 
\begin{equation}\label{eq:evomulti}
\begin{split}
&\dot{\varphi_i}=-\frac{M_i}{\varepsilon}\bigg[\frac{\delta F_{\rm multi}}{\delta \varphi_i}+\lambda\sqrt{2H({\varphi_i})}\bigg],\\
&\lambda=\frac{\sum_{j=1}^N M_j \frac{\delta F_{\rm multi}}{\delta \varphi_j}}{\sum_{j=1}^N M_j \sqrt{2H({\varphi_j})}}.
\end{split}
\end{equation}
with $M_i\equiv M(\hat{\mathbf{n}}_i)$. $\lambda$ may be chosen with different forms\cite{Brassel2011,BRETIN2018324,bretin2019phase}. 
This approach has some similarity with other, well-known, multi-phase field approaches\cite{Chen1994,steinbach1996phase,Moelans2008,Steinbach_2009,DARVISHIKAMACHALI20122719,Toth2015,DIMOKRATI2020147}. 
However, it allows to directly control of interface properties through a convenient parametrization that allows us to recover the general sharp-interface dynamics (as outlined in Sect.~\ref{sec:cmodel}) and ensures other useful features such as an orientation-independent interface thickness \cite{Tor2009}.

At variance with the model presented in Sect.~\ref{sec:PF}, Eqs.~\eqref{eq:Fmulti} and \eqref{eq:evomulti} have parameters, such as $M$ and $\gamma$, associated with phases rather than the interfaces between them. 
However, for isotropic $M_i$ and $\gamma_i$, these dynamics reproduce generalised mean-curvature flow of the interface between phases $i$ and $j$ with 
$M_{ij}^{-1}\dot{\mathbf{x}}=\gamma_{ij} \kappa \hat{\mathbf{n}}$ 
(for $\varepsilon \rightarrow 0$)\cite{BRETIN2018324} and similar results are expected in an anisotropic setting \cite{garcke1998anisotropic,garcke1999multiphase,Ratz2006,Tor2009}. Interface properties may be related to parameters entering Eq.~\eqref{eq:evomulti} as $\gamma_{ij}=(\gamma_i+\gamma_j)/2$ and $M_{ij}^{-1}=M_i^{-1}+M_j^{-1}$. 
Targeted $M_{ij}$ and $\gamma_{ij}$ may then be set through suitable definitions of $M_i$ and $\gamma_i$. For instance, focusing on the latter, one may  compute
\begin{equation}
    \gamma_i=\frac{\sum_{j}^N \gamma_{ij} |\nabla \varphi_i| |\nabla \varphi_j|}{\sum_{j}^N |\nabla \varphi_i| |\nabla \varphi_j| + \delta}, \qquad \text{with }\quad  \gamma_{ii}=0. 
    \label{eq:gamma_par}
\end{equation}
This enforces $\gamma_i=\gamma_j=\gamma_{ij}$ at the $ij$-interfaces and an average of the properties of the  $ij$-interfaces meeting at triple junctions. 
$\delta=10^{-6}$ is set to avoid numerical divergences away from interfaces, similarly to Eq.~\eqref{eq:levelset}. 
Notice that, by extension, one may exploit a form as in \eqref{eq:gamma_par} with the product of three $|\nabla \varphi_i|$ terms to enforce properties for triple junctions only. $M_{ij}$ can be set similarly. 
As a practical example we may consider a microstructure having grain with different orientations. 
To set $\gamma_{ij}$ accordingly, one may then exploit Eqs.~\eqref{eq:gammaaniso} for an inclination angle 
\begin{equation}
\phi = \arctan\bigg(\frac{\hat{n}_2}{\hat{n}_1}\bigg) - \frac{\theta_i + \theta_j}{2},
\label{eq:phigij}
\end{equation}
where $\theta_{i,j}$ are the orientations assigned to domains $i,j$ and the inclination of the reference interface is $(\theta_i + \theta_j)/2$. $\gamma_{i}$ can then be defined through Eq.\eqref{eq:gamma_par}. A corresponding numerical example is shown in Sect.~\ref{sec:num}.

To include the contribution of stress fields generated by disconnection Burgers vectors and the external stress and  to account for differences in chemical potentials between different phases (as in Sect.~\ref{sec:PF}), we include an advection-like term accounting for (multiple) distinct interfaces/phases. 
The net velocity of an interface must be the sum of the normal velocities from the phases meeting at that interface point. 
For the evolution of $\varphi_i$, describing the $i-th$ phase, i.e. in $\dot{\varphi}_i$, we include
\begin{equation} 
\Phi_i=\sum_{i=j}^{N} \bigg( (\bar{\tau}_{\rm self}+\tau_{\rm ext})\Lambda_{ij}' + \mu_j\bigg) |\nabla \varphi_j|(\hat{\mathbf{n}}_i \cdot \hat{\mathbf{n}}_j).
\label{eq:advectionmulti}
\end{equation}
The last product accounts for the relative directions of $\hat{\textbf{n}}_j$ with respect to $\hat{\textbf{n}}_i$. $\mu_j$ is the chemical potential of the $j^\text{th}$ phase. $\Lambda_{ij}'=\varepsilon_{ij}(\beta^{(2)}_{ij}-\beta^{(1)}_{ij})$, with $\beta^{(k)}_{ij}=\beta^{(k)}_{ji}$  ($\varepsilon_{ij}$ the Levi-Civita symbol), accounts for the dislocation character of the disconnections at the interface between phases $i$ and $j$. 

We may understand Eq.~\eqref{eq:advectionmulti} by considering an interface between two phases (1,2), i.e. $\varphi_1=1-\varphi_2$  (see e.g. Fig.~\ref{fig:figure3}b at $x\sim0.7$). 
Here,  $\hat{\mathbf{n}}_1=-\hat{\mathbf{n}}_2$, and both $\nabla \varphi_1=-\nabla \varphi_2$, $|\nabla \varphi_1|$ = $|\nabla \varphi_2|$. 
With these relations, Eq.~\eqref{eq:advectionmulti} yields
\begin{equation}
\begin{split}
\Phi_1
&=(\bar{\tau}_{\rm self}+\tau_{\rm ext})\underbrace{(\beta_{12}^{(2)}-\beta_{12}^{(1)})}_{\Lambda_{12}}|\nabla \varphi_1| + \underbrace{(\mu_1-\mu_2)}_{\psi_{12}} |\nabla \varphi_1|,\\
\Phi_2
&=-(\bar{\tau}_{\rm self}+\tau_{\rm ext})\Lambda_{12}|\nabla \varphi_1| -\psi_{12} |\nabla \varphi_1|.
\end{split}
\end{equation}
If $\beta^{(k)}_{12}=0$ and $\mu_1=\mu_2$, one trivially finds $\Phi_1=\Phi_2=0$ (i.e. no advection occurs in the absence of  dislocation character and interfacial chemical potential jumps). 
If $\mu_1\neq \mu_2$ and/or 
$\beta^{(1)}_{12}\neq \beta^{(2)}_{12}$, we find $\Phi_1=-\Phi_2$ and the advection of the two phases occurs  in the same direction with velocity $|\nabla \varphi_1| (\psi_{12} + (\bar{\tau}_{\rm self}+\tau_{\rm ext})\Lambda_{12})$. 

With many interfaces, the self-stress $\tau_{\rm self}(s)$ at any interface point $s$  is found by integration over all interfaces in the system; i.e.,
\begin{equation}
\tau_\text{self}(s)
= \sum_{j>i}^N 
\left(\beta^{(1)}_{ij} I^{(1)}_{\Sigma_{ij}}(s) + \beta^{(2)}_{ij} I^{(2)}_{\Sigma_{ij}}\right), 
\end{equation}
where $\Sigma_{ij}=\Sigma_{i}\cap\Sigma_{j}$ corresponds to an interpolation of the 0.5 level sets for $\varphi_i$ and $\varphi_j$ (that reduces to $\varphi_i\varphi_j\sim 0.25$ at interfaces among two phases). 
When approaching triple junctions, the interpolation of the two closest phases (having the largest product $\varphi_i\varphi_j$) may be considered. 
This realizes a triple junction as a point in the sharp-interface limit only, while delivering a description compatible with the diffuse interface approach otherwise. 
Note that setting $\beta_{ij}$ implies a choice of the  interface normal or tangent vector orientation. 
For the single interface approach of Sect.~\ref{sec:PF}, this is inherently defined as the normal pointing  towards the $\varphi=0$ phase. 
The extension of $\tau(s)$ along the interface normal(s) is performed as in Eq.~\eqref{eq:levelset}. 
An example of $\tau_{\rm self}(\mathbf{r})$ in a microstructure is reported in Sect.~\ref{sec:num}. 

\begin{figure*}
   \includegraphics[width=\textwidth]{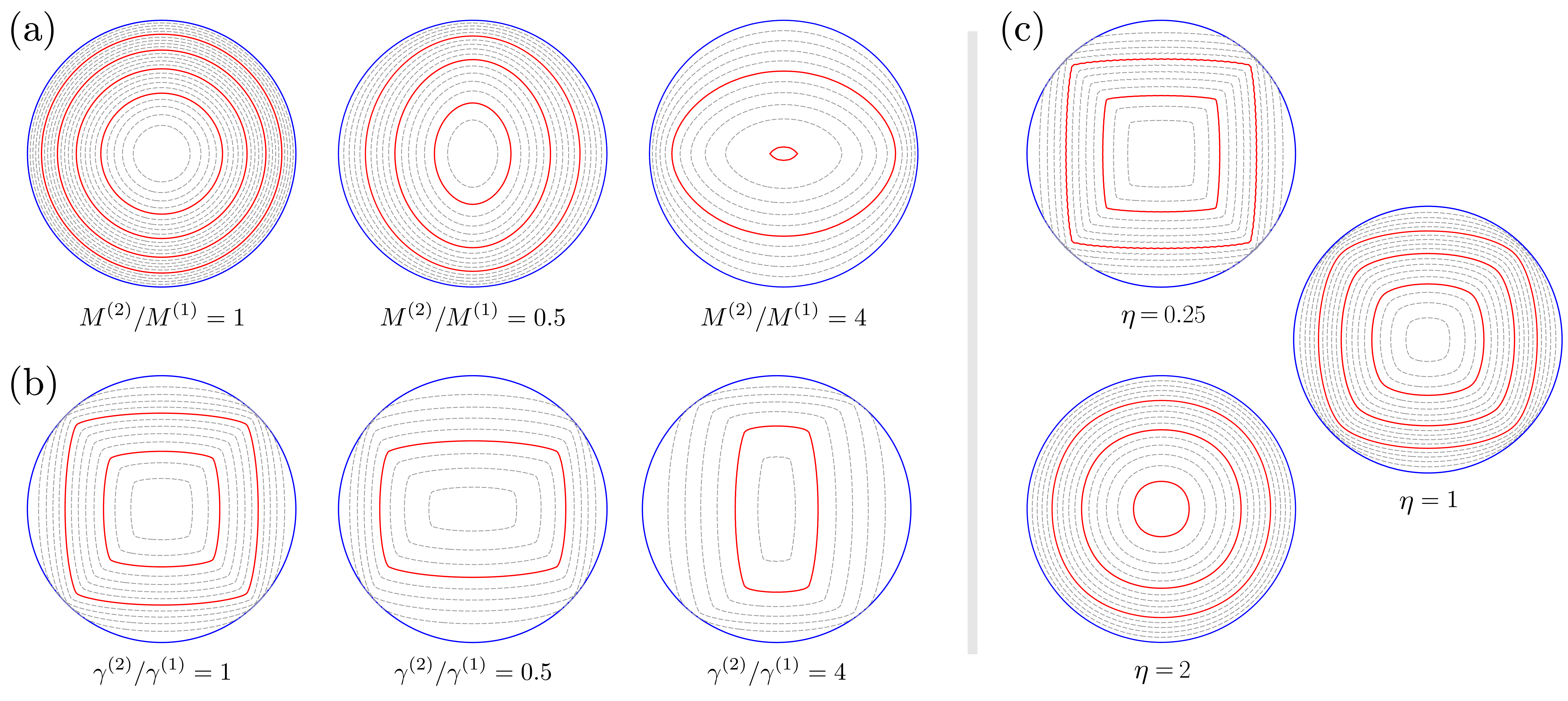} 
   \caption{Shrinking of an initially circular (radius $\tilde{R}=100$) embedded domain and the effect of interface energy and/or mobility anisotropies (parameters not specified here may be found in Sec.~\ref{sec:parnum}). (a) Effect of the $M^{(2)}/M^{(1)}$ ratio with isotropic interface energy density. (b) Effect of the $\gamma^{(2)}/\gamma^{(1)}$ ratio with isotropic interface mobility ($\eta=0.5$). (c) Effect of the regularisation parameter $\eta$ (see Eq.~\eqref{eq:gammaaniso_R}) with $\gamma^{(2)}/\gamma^{(1)}=1$.
   The solid blue  curves show the initial two-phase interface configuration and the solid red and  dashed grey curves show the interface position at different times, with  time increments  $\Delta \tilde{t}_{\rm plot}=200$.}
    \label{fig:figure4}
\end{figure*}

The final, multi-phase field, general evolution equation, is 
\begin{equation}
\begin{split}
\dot{\varphi_i}=&\frac{M_i}{\varepsilon} \bigg[ \varepsilon\Phi_i-\frac{\gamma_i}{\varepsilon}H'(\varphi_i)+\lambda\sqrt{2H({\varphi_i})}\\
&
+\varepsilon  \nabla \cdot \bigg( \gamma_i \nabla \varphi_i +
|\nabla \varphi_i|^2 \mathbf{P}_i \nabla_{\hat{\mathbf{n}}_i} \gamma_i\bigg)\bigg].
\label{eq:acmulti}
\end{split}
\end{equation}
with $(1/\varepsilon)H(\varphi_i)\stackrel{\varepsilon \rightarrow 0 }{\approx} (\varepsilon/2)|\nabla \varphi_i|^2$ and additional definitions as in Eqs.~\eqref{eq:ac} and \eqref{eq:evomulti}.

\section{Numerical Results and Discussion}
\label{sec:num}

In this section, we show the effects of different driving forces on disconnection-mediated interface evolution as encoded in the diffuse interface through full simulations and proof of concepts.

\subsection{Parameters and Simulation Details}
\label{sec:parnum}

The numerical results illustrated in this section are obtained by numerical integration of the disconnection-dynamics-based phase-field partial differential equations \eqref{eq:ac} and \eqref{eq:acmulti}. 
We exploit a semi-implicit integration scheme and approximate the non-linear terms using a single-iteration Newton method. 
The discretisation may exploit both Finite Difference and Finite Element Methods\cite{Vey2007,WitkowskiACM2015}. 
We employ periodic boundary conditions (PBCs) for the evolution law of $\varphi_i$.
We also determine $\tau_{\rm self}$ by summing the elastic field of image interfaces repeated periodically (computational domain periodicity). 
Additional details are provided in the Supplemental Material.

\begin{figure*}
   \includegraphics[width=1\textwidth]{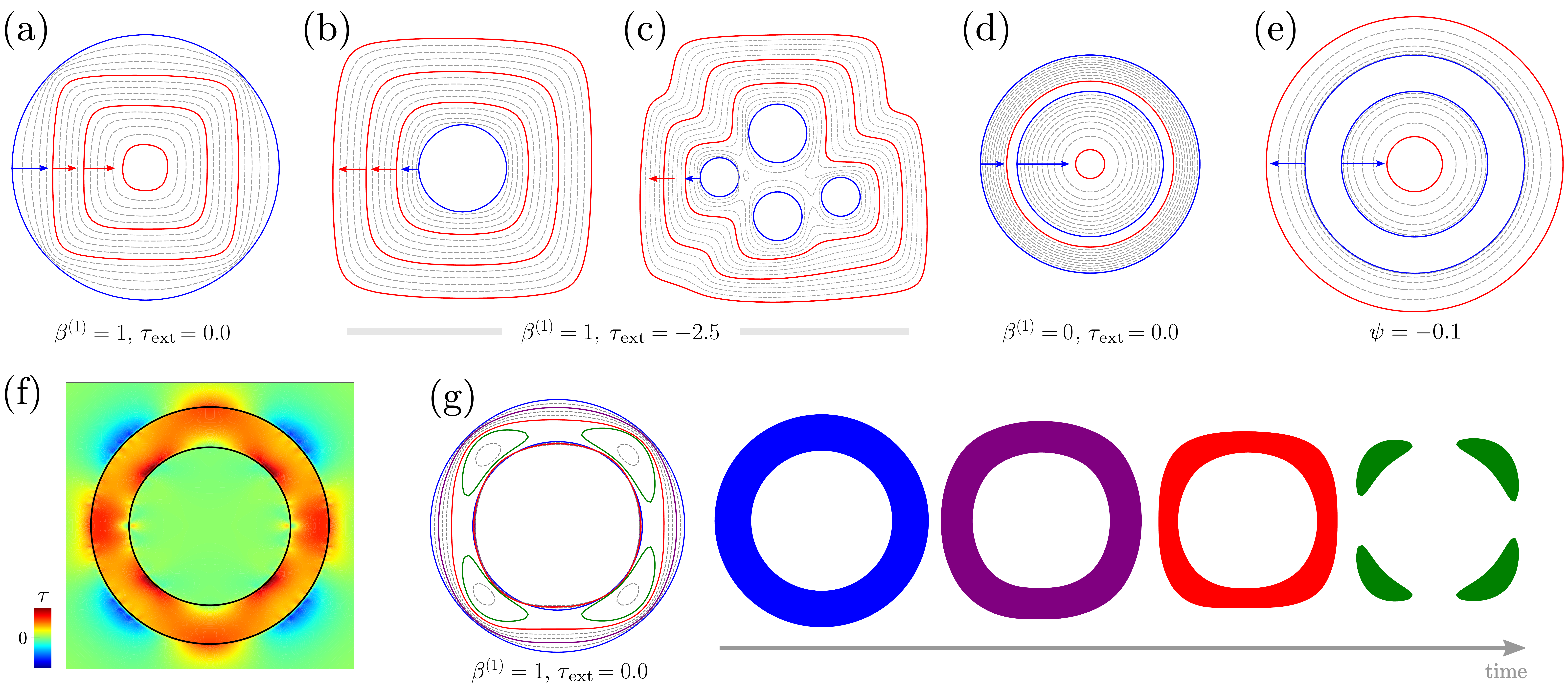} 
   \caption{Effects of disconnection Burgers vector and stress field (parameters not specified here may be found in Sec.~\ref{sec:parnum}). (a) Shrinkage of a circular interface with initial radius $\tilde{R}=100$ including the effect of dislocation self-stress ($\beta^{(1)}=1$, $\tau_{\rm ext}=0$, $\Delta \tilde{t}_{\rm plot}=100$). (b) Effect of external stress field on a circular interface as in  (a): $\beta^{(1)}=1$, $\tau_{\rm ext}=-2.5$, $\Delta \tilde{t}_{\rm plot}=1250$ (same scale as  (a)). (c) Same as (b) but for multiple inclusions which coalesce  (same scale as (a)), (d) Annular $\varphi=1$ domain shrinking by mean curvature flow (initial inner and outer radii are $\tilde{R}=100$ and $\tilde{R}=150$, $\Delta \tilde{t}_{\rm plot}=500$). (e) The same annular domain evolving with $\psi = -0.1$, $\Delta \tilde{t}_{\rm plot}=80$. (f) Self-stress generated by the two annulus interfaces for $\beta^{(1)}=1$ and $\tau_{\rm ext}=0$. (g) Evolution of the annulus under the action of the self-stress as in panel (f) - the coloured images are the evolving annulus shape correspond to the same colour contours in the first image.}
    \label{fig:figure5}
\end{figure*}

Our standard setup (against which others are compared) corresponds to classical mean curvature flow (i.e. fully isotropic interface energy and mobility with no stress or chemical potential variation: $\beta^{(1)}_{ij}=\beta^{(2)}_{ij}=0$, $\tau_{\rm ext}=0$, $\psi=0$ or $\mu_i=0$, $\gamma(\hat{\mathbf{n}}_i)=\gamma_0=1$, $M^{(1)}_i=M^{(2)}_i=M_0=1$, $a=\alpha$, and $G/[2\pi(1-\nu)] = 1$). 
In what follows, we only explicitly describe changes from this basic case. 
For the geometries and sizes considered in the following, we found that an interface thickness of $\varepsilon=10\alpha$ (or $\tilde{\varepsilon}=10$)  satisfactorily reproduces the sharp interface limit (see, in particular, simulations and discussions of Figs.~\ref{fig:figure4} and \ref{fig:figure5}).

\subsection{Single interface}

A few examples of the evolution of an initially circular domain (initial radius $\tilde{R}=100$) are shown in Fig.~\ref{fig:figure4}; these illustrate the influence of interface energy and mobility anisotropy (in the absence of stress or chemical potential jumps). 
These examples show interface evolution (0.5 isolines of $\varphi$); solid red and grey dashed curves show the interface position at uniform (reduced) time intervals, $\Delta \tilde{t}_{\rm plot}=200$.
Figure~\ref{fig:figure4}a shows the effect of interface mobility anisotropy ($M^{(2)}/M^{(1)}$) for isotropic interface energy. 
For  isotropic mobility $M^{(2)}/M^{(1)}=1$ (i.e.,  classic mean curvature flow) the domain shrinks as a circle and disappears at  $\tilde{t}_{\rm end}=R^2/2M_0=5000$. In practice, effective numerical convergence is achieved with the considered $\varepsilon$.
Varying the disconnection mobility ratio $M^{(2)}/M^{(1)}$ (Eq.~\eqref{eq:modelM}) leads to the evolution of the initially circular domain into ellipses with increasing ellipticity as the domain shrinks.  
Changing $M^{(2)}/M^{(1)}$ from $>1$ to $<1$ rotates the major axis  of the ellipse. 
Figure~\ref{fig:figure4}b shows the effect of anisotropic interface energy densities (with $\eta=0.5$) and isotropic mobility. 
While the equilibrium domain shape is determined by the interface energy anisotropy (Wulff shape \cite{Wulff1901,Herring1951}),
the domain shape varies with time for different  $\gamma^{(2)}/\gamma^{(1)}$ ratios. 
Fixing the ratio $\gamma^{(2)}/\gamma^{(1)}=1$ and varying the regularisation parameter $\eta$ (see Eq.~\eqref{regular}) in  Fig.~\ref{fig:figure4}c, we see that $\eta$ controls facet flatness; large $\eta$ leads to  isotropic shapes, while $\eta\to0$ gives flat facets with converging morphologies and time scales. 
Figure~\ref{fig:figure4} provides a means of comparison of the phase-field model predictions with their sharp-interface counterparts\cite{Han2021} and, generally, by anisotropic mean curvature flow\cite{taylor1998diffuse,garcke1999anisotropy,stocker2007effect,Li2009}.

Figure~\ref{fig:figure5} illustrates the role played by the disconnection Burgers vector in coupling the evolutions of the interface to the external and internal (self-stress) fields. In particular, Fig.~\ref{fig:figure5}a illustrates the dynamics obtained with $\beta^{(1)}=1$ (i.e., $\mathbf{b}=b^{(1)}\mathbf{e}_1$,  $b^{(1)}=h^{(1)}$). 
The initially circular domain shrinks (as expected on the basis of mean curvature flow), but with a near-square shape because of the self-stress $\tau_{\rm self}$ and disconnection glide along the $\mathbf{e}_1$ and $\mathbf{e}_2$ directions. 
This anisotropic interface evolution occurs with isotropic interface energy and mobility and is a consequence of the dislocation character of disconnections. 
The same dynamics is achieved with $\beta_1=-1$, i.e. $\mathbf{b}=-b^{(1)}\mathbf{e}_1$.
However, the finite Burgers vectors make no contribution to the shape evolution when  $\beta_1=\beta_2$, but double the interface velocity for $\beta_2=-\beta_1$. 
The evolution shown in Fig.~\ref{fig:figure5}a matches the evolution for the same initial shape and parameters obtained by the sharp-interface approach (see Part I\cite{Han2021}); this validates the evaluation of stresses, the translation into the considered PF approach, and the numerical convergence achieved with the considered $\varepsilon$  when stresses are present. 
Figure~\ref{fig:figure5}b shows the effect of an external shear stress ($\mathbf{b}$ as in Fig.~\ref{fig:figure5}a);  
for this applied stress ($\tau_{\rm ext}=-2.5$), the initially circular domain grows rather than shrinks, as would occur by mean curvature flow alone, and develops a four-fold shape (qualitatively resembling that in Fig.~\ref{fig:figure5}a), still with isotropic interface energy and mobilities. Figure~\ref{fig:figure5}c illustrates the same applied stress and $\beta$ as in Fig.~\ref{fig:figure5}b, but for four, initially circular domains.
Here, we see that these domains grow, impinge and merge, demonstrating the ability of this method to naturally accommodate topology changes rather than just shape evolution. 
In the long time limit, this microstructure evolves into one identical with Fig.~\ref{fig:figure5}b. 

Figures~\ref{fig:figure5}d--\ref{fig:figure5}g show the temporal evolution of an annular domain (the classical mean curvature flow limit is illustrated in \ref{fig:figure5}d). 
Since the inner interface has a smaller radius (larger curvature) than the outer radius, the annulus thickens as it shrinks, eventually becoming a circle prior to disappearing. 
The addition of a chemical potential jump that favours the growth of the annular ($\varphi=1$) domain at the expense of the inner and outer domains (see Fig.~\ref{fig:figure5}e), leads to an increase in annulus area (the inner radius shrinks and the outer radius grows).

\begin{figure}[t]
   \includegraphics[width=1\columnwidth]{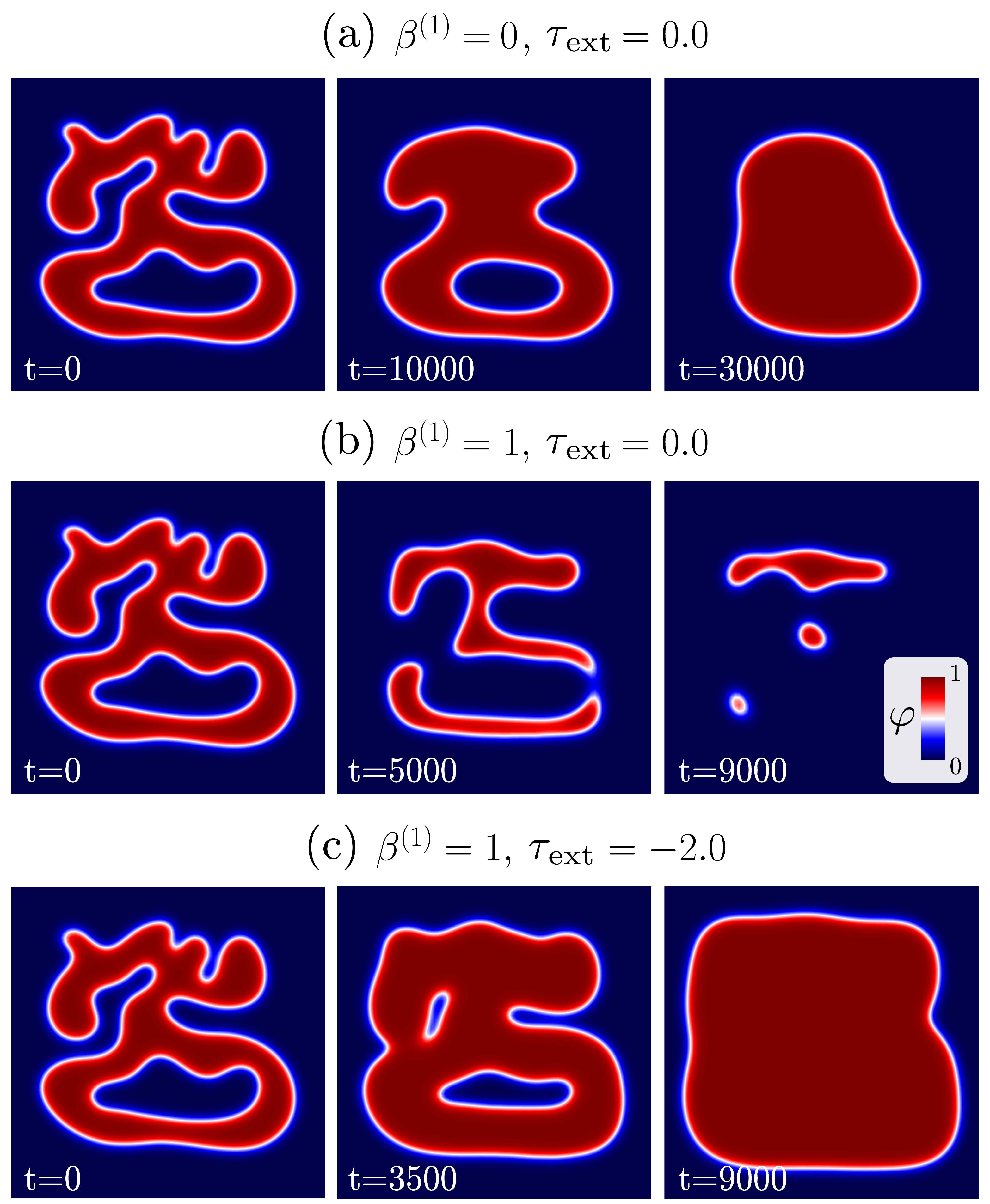} 
   \caption{Effects of disconnection Burgers vector and applied stress on the evolution for the same initial, complex domain (parameters not specified here may be found in Sec.~\ref{sec:parnum}). (a) classical mean curvature flow dynamics and for a finite Burgers vector ($\beta^{(1)}=1$) (b) without  and (c) with an external stress $\tau_{\rm ext}=-2.0$. }
    \label{fig:figure6}
\end{figure}

When the disconnections have non-zero Burgers vectors ($\beta^{(1)}=1$, $\tau_{\rm ext}=0$, $\psi=0$, isotropic interface energy and mobility), the inner and outer interfaces move toward each other with velocities varying along each interface (Figs.~\ref{fig:figure5}f and \ref{fig:figure5}g). 
This is a result of the self-stress developed as the interfaces move (i.e., shear coupling)  - see Fig.~\ref{fig:figure5}f.
Here, the annulus shrinks, and thins non-uniformly,  pinching off into a four domains - see Fig.~\ref{fig:figure5}g. 
A nearly stationary inner interface is observed; this is different from the evolution for the same interface(s) with step character alone (see Fig.~\ref{fig:figure5}d). 
This non-trivial evolution, strongly deviates from mean curvature flow, results from the self-stress that accompanies disconnection glide in orthogonal directions and topology change that is only observable with the disconnection-based, shear-coupled interface migration model (accommodating disparate driving forces).

The behaviours depicted in Fig.~\ref{fig:figure5} for initially circular domains are also observed for arbitrary domain geometries and topologies as shown for a non-trivial initial domain shape in Fig.~\ref{fig:figure6}. 
Evolution of the same complex domain by mean curvature flow, the disappearance of the $\varphi=1$ phase due to the action of the self-stress field and the growth of the $\varphi=1$ phase with a four-fold symmetric interface shape for a large, negative external stress are all illustrated in the three sets of panels in Fig.~\ref{fig:figure6}. 
Again, internal and external stress effects, disconnection glide in orthogonal directions, and the competition between driving forces all strongly modify domain morphology evolution.
We note that in all three complex cases, the domains evolve toward convex morphologies but, when topological change occurs, this may result in multiple convex domains.
The shrinking and disappearing of arbitrarily complex domain shapes as convex objects is well-known in mean curvature flow\cite{gage1986heat,grayson1987heat}, but no topological changes are expected. This emerges from consideration of stresses acting on disconnection at interfaces with non-zero Burgers vector.

\subsection{Multiple domains}

\begin{figure}[t!]
   \includegraphics[width=\columnwidth]{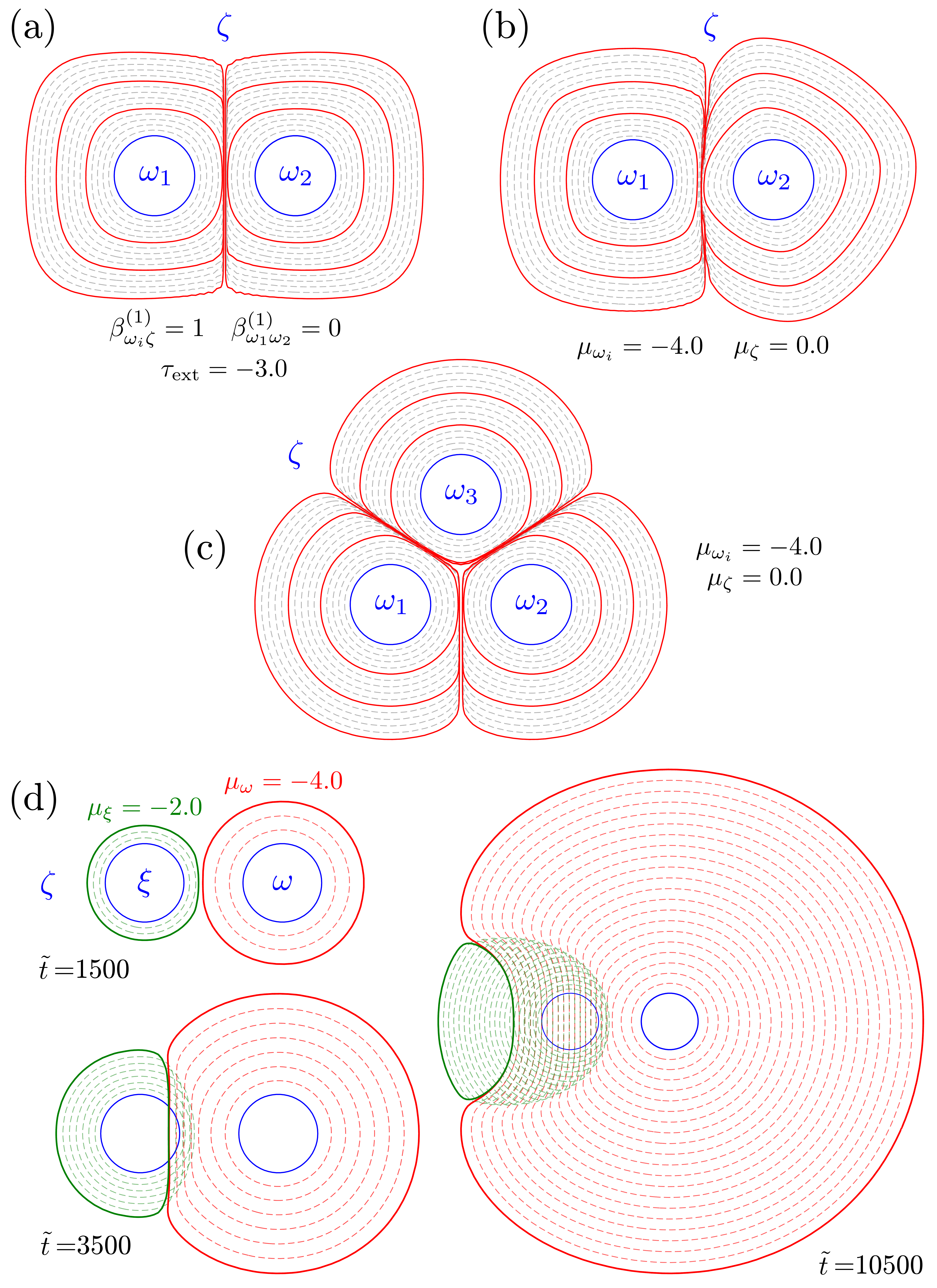} 
   \caption{Growth of multiple domains (inclusions) of phase $\omega_i$ in $\zeta$. 
   Initial $\omega_i$ domains  (blue circles) have radius $\tilde{R}=100$ (parameters not specified here may be found in Sec.~\ref{sec:parnum}).
   (a) Growth and impingement of two domains $\omega_i$ with an external shear stress ($\beta^{(1)}_{\omega_i\zeta}=1$, $\beta^{(1)}_{\omega_1\omega_2}=0$, $\beta^{(2)}_{ij}=0$, $\tau_{\rm ext}=-3.0$, $\Delta \tilde{t}_{\rm plot}=1250$). (b) Growth of two domains with $\mu_{\omega_i}<\mu_{\zeta}$ with  interface-energy anisotropies rotated by $\omega_i \zeta$ ($\Delta \tilde{t}_{\rm plot}=200$). (c) Growth of three phases $\omega_i$ with $\mu_{\omega_i}<\mu_{\zeta}$ forming a triple junction ($\Delta \tilde{t}_{\rm plot}=200$). (d) Growth of two phases with different chemical potentials, $\mu_\omega < \mu_\xi < \mu_\zeta=0$.} 
 \label{fig:figure7}
\end{figure}

\begin{figure*}
   \includegraphics[width=\textwidth]{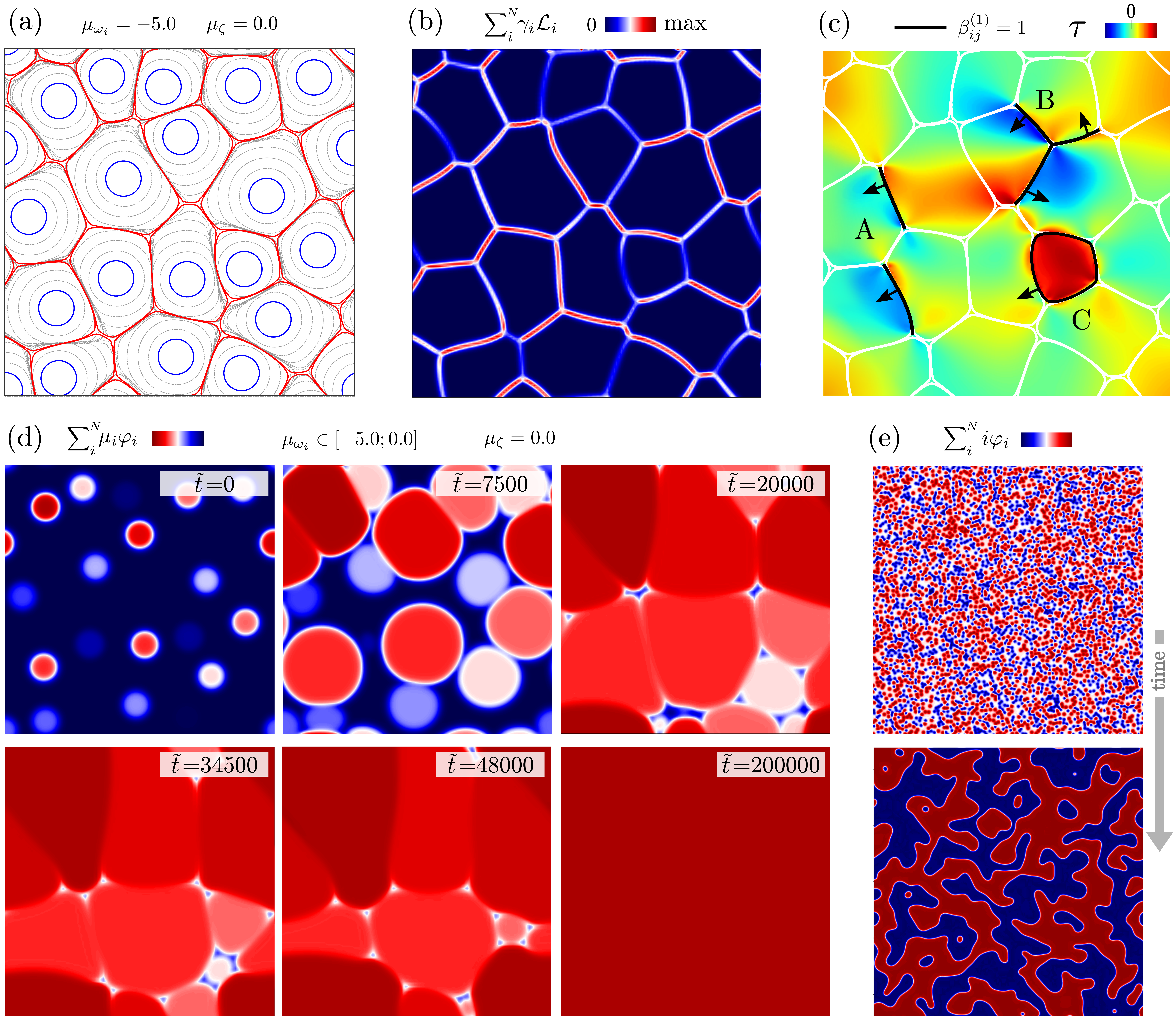} 
\caption{Examples of systems with multiple interfaces (parameters not specified here may be found in Sec.~\ref{sec:parnum}). 
(a) Isotropic growth of 20 inclusions, represented by different phases $\omega_i$ (initially  circular inclusion of $\tilde{R}=150$), in $\zeta$ leading to the formation of a polycrystalline microstructure ($\mu_{\omega_i}=-5.0$, $\mu_{\zeta}=0.0$, $\Delta \tilde{t}_{\rm plot}=1500$). 
(b) Anisotropic interface energy density, with $\mathcal{L}_i=(1/2)\varepsilon |\nabla \varphi_i|^2+(1/\varepsilon)H(\varphi_i)$, for a microstructure corresponding to the last stage in panel (a) and randomly assigned grain rotations. Anisotropic interface energy is set as $\gamma_{ij}(\phi)$ via Eqs.~\eqref{eq:gammaaniso_R}, \eqref{eq:phigij}, \eqref{eq:gamma_par}, with $\gamma^{(1)}=1$ and $\gamma^{(2)}=4$. 
(c) Self-stress for a microstructure corresponding to the last stage in panel (a) and a prescribed distribution of $\beta_{ij}$ illustrating different cases for interfaces with non-vanishing dislocation character: isolated interfaces (A), a triple junction (B), a closed interface (C). Black lines correspond to interfaces with $\beta_{ij}^{(1)}=1$, while arrows illustrates their assigned normal vector.
(d) Growth of the inclusions from (a) with random $\mu_{\omega_i} \in [-5.0,0]$ for different inclusions ($\mu_\zeta=0$). (e) Large number of inclusions ($\sim4000$) of  two phases with $\mu_{\omega_i}=-5.0$ ($\mu_\zeta=0$).}
    \label{fig:figure8}
\end{figure*}

We now consider the case of multiple domains, where interface junctions occur. First consider the growth of multiple domains (particles) of phase $\omega_i$ growing in $\zeta$, as illustrated in Fig.~\ref{fig:figure7}a-\ref{fig:figure7}c.
Figure~\ref{fig:figure7}a shows the case of two, initially circular, isolated particles growing together under the influence of an external stress $\tau_\text{ext}$, where each $\omega_i\zeta$ interface has the same, finite $\beta_{\omega_i\zeta}$ but where the $\omega_1\omega_2$ interface has $\beta_{\omega_1\omega_2}=0$.
This case is similar to Fig.~\ref{fig:figure5}b prior to impingement.
These results demonstrate how two $\omega_i\zeta$ interfaces smoothly merge to form a $\omega_1\omega_2$ interface and two triple (three domain) junctions $\omega_1\omega_2\zeta$.

Figure~\ref{fig:figure7}b shows the growth of the two $\omega_i$ domains into the $\zeta$ phase for the case where the chemical potentials of the two phases are different, i.e.,  $\mu_{\omega_i}<\mu_\zeta$.
In this example, the two $\omega$ phase domains have different crystallographic orientations and hence rotated $\omega\zeta$ interface energy anisotropies (by $\pi/5$). 
This demonstrates the impingement and growth of identical but rotated crystals (as in grain growth).
Figure~\ref{fig:figure7}c shows the growth of three, differently oriented $\omega$ grains to form a classic 3-grain triple junction, with  triple junction angle $2\pi/3$ for the equal grain boundary energy case $\gamma_{\omega_1\omega_2}=\gamma_{\omega_2\omega_3}=\gamma_{\omega_3\omega_1}$.

Finally, Fig.~\ref{fig:figure7}d shows the case of two dissimilar phases ($\omega$ and $\xi$) growing from $\zeta$, with $\mu_\omega < \mu_\xi$.
Such three-phase, single-component co-existence can occur at a fixed temperature  $T$ and pressure $p$ or along a curve in $T$-$p$ space during the kinetic disappearance of $\zeta$ or, in a finite $T$-$p$ parameter range in the presence of a magnetic field (as per the Gibbs phase rule). 
This example shows the growth of two domains at different rates, the formation of a three-phase interface, and the translation of the centre of mass of one phase relative to the others (green outlined phase in Fig.~\ref{fig:figure7}d).

The generalisation of our approach to a large number of interfaces is shown in Fig.~\ref{fig:figure8}, and demonstrates that this approach is applicable to complex microstructures rather than individual interfaces. 
Fig.~\ref{fig:figure8}a first illustrates the growth of many $\omega_i$ crystalline particles into $\zeta$ resulting in a polycrystalline microstructure. Here, there is no external stress, yet internal stress ($\tau_{\rm self}$) may develop as a result of disconnection 
motion during interface/grain-boundary migration. 
Two examples of disconnection-related properties associated to this resulting microstructure are shown in Figs.~\ref{fig:figure8}b and \ref{fig:figure8}c. 
In particular, \ref{fig:figure8}b shows the anisotropic interface energy density  corresponding to the final stage shown in \ref{fig:figure8}a for randomly oriented grains with $\gamma_{ij}$ set through Eqs.~\eqref{eq:gammaaniso_R}, \eqref{eq:phigij}, and \eqref{eq:gamma_par}. 
Fig.~\ref{fig:figure8}c illustrates $\tau_{\rm self}$ for arbitrarily assigned $\beta^{(1)}_{ij}=0$ (white lines) and $\beta^{(1)}_{ij}=1$ (black lines) to interfaces in the microstructure obtained in Fig.~\ref{fig:figure8}a. 
Different cases  dislocation character cases are illustrated here: isolated surfaces (A), a triple junction (B), and a closed interface (C). Notice that the latter reproduces a stress distribution qualitatively corresponding to Fig.~\ref{fig:figure1}e. 
This example demonstrates that the description of microstructures in the presence of accumulation/release of stress, as seen in molecular dynamics simulations~\cite{thomas2017reconciling}, can be described by the  approach presented here.

Figure~\ref{fig:figure8}d  shows the evolution of a microstructure of several phases $\omega_i$ with different $\mu_{\omega_i}$ (this is a generalisation of the cases shown in Figs.~\ref{fig:figure7}d and \ref{fig:figure8}a). 
The colour map indicates the chemical potential of the phases.
This example may correspond, for instance, to the growth and coarsening of a polycrystal when applying magnetic fields, thus enforcing preferential orientations \cite{BackofenPRL2019}. 
In this simulation, the lower $\mu_{\omega_i}$ the faster the phase grows, both before and after forming a dense microstructure with $\omega_i$ phases only. 
The coarsening of the phases with lowest $\mu_{\omega_i}$ occurs, eventually filling the entire domain (see the dark red colour, corresponding to the lowest $\mu_{\omega_i}$).
Note that the three-domain junctions move according to jumps in the chemical potential across all of the different interfaces.

Finally, Fig.~\ref{fig:figure8}e illustrates an example of a large system initialised with $\sim 4000$ small particles of  two different $\omega_i$ phases with  $\mu_{\omega_1}=\mu_{\omega_2}=-5.0$ growing into a parent phase  $\mu_\zeta=0$.
In this case, the resultant domain morphology is a classical mazed (or Ising) microstructure, which was, for example, observed in the growth of a Au thin film  on a $\{0 0 1\}$ Ge substrate~\cite{radetic2012mechanism}.
Here, each phase grows, but interfaces do not intersect after the parent $\zeta$ phase disappears. 

\section{Conclusions}
\label{sec:conclusions}

A general continuum framework for simulating the evolution of a  microstructure, consisting of an interface network separating crystalline domains (e.g., grain boundaries in a polycrystal or heterophase interfaces in a multiphase microstructure) is presented. 
The framework is based upon a disconnection mechanism-based description of interface migration.
The simulation method accounts for a wide range of driving forces for microstructural evolution, including chemical-potential differences between domains, capillarity (interface energy/curvature), external stresses, as well as the stresses generated by microstructure evolution itself.
The model also includes anisotropy in both thermodynamic driving forces as well as kinetic coefficients (mobilities).
The continuum approach is based upon the phase-field method and, as such, naturally accommodates both complex microstructures, topology change and the interplay of different physical effects.

Selected numerical simulations illustrate the diverse and robust capabilities of the approach.
In particular, we show examples that illustrate the effects of anisotropies in both mobility and interface energy density, applied stresses, and differences in energies between competing phases in simple and complex microstructures. 
These simulations also clearly show how the microscopic, underlying disconnection mechanism of interface motion gives rise to effects 
seen in atomic-scale (molecular dynamics) simulations and experiments. The emerging dynamics deviates from mean curvature flow, leading to additional anisotropies, topological changes and grain migration.
Of particular note is the inclusion of shear coupling and its constraint and accommodation in polycrystalline microstructure evolution. While general features are presented here, future works will be devoted to detailed investigations of these effects on grain growth and evolution of microstructures.

The work presented in this two-paper study delivers a versatile framework for the evolution of polycrystalline and multiphase microstructures based upon the underlying mechanisms of interface migration. 
This approach may be further extended. 
For example,  it may be generalised to account for multiple disconnection types on each reference system and for non-orthogonal reference systems \cite{Han2021,Zhang2018}.
In addition, the chemical potential and external fields  (e.g., electromagnetic field) may be incorporated and allowed to vary throughout space.  
The effects of misfit between particles and the matrix, or in general different grains, may be included by explicitly accounting for the solution of the elastic problem following, e.g., Ref.~\onlinecite{Ratz2006,Salvalaglio2018} and providing its generalization to systems with many interfaces. 
Composition fields and composition-dependent material parameters may also be included; this requires the adoption of conserved dynamics for some of the variables in addition or in pace of the present overdamped interface dynamics.
While we focus on two-dimensional examples in the present work, the extension to three dimensions is both natural and straightforward within the diffuse interface description of evolving interfaces. 
However, the generalisation of the underlying disconnection model to two-dimensional interfaces in three dimensions requires the inclusion of several features not discussed here; e.g., curved disconnections (spatially varying line directions) may be described using the same ideas as are inherent in dislocation dynamics.  
Finally, while the numerical methods described here for solution of the dynamical evolution equations were kept simple to facilitate transparent explanations, these may be greatly improved for systematic, large-scale simulations (e.g., through advanced adaptive finite element methods \cite{Backofen2019,Praetorius2019}).

\vspace{30pt}
\section*{Acknowledgements}
MS acknowledges support from Visiting Junior Fellowship of the Hong Kong Institute for Advanced Studies and the Emmy Noether Programme of the German Research Foundation (DFG) under Grant SA4032/2-1. 
DJS  acknowledges  support from the Hong Kong Research Grants Council Collaborative Research Fund C1005-19G. 
JH acknowledges support from City University of Hong Kong Start-up Grant 7200667 and Strategic Research Grant (SRG-Fd) 7005466.
We gratefully acknowledge computing time grants from the Centre for Information Services and High Performance Computing (ZIH) at TU Dresden.

%

{\textcolor{white}{
\widetext
}}
\newpage
\onecolumngrid
\begin{center}
\textbf{\large {\sc{supplementary material}}\\
Disconnection-Mediated Migration of Interfaces in Microstructures:\\
II. diffuse interface simulations}\\[.3cm]
Marco Salvalaglio$^{1,2,3}$, David J. Srolovitz$^{4}$, Jian Han,$^{5}$\\[.1cm]
  {\itshape \small
  ${}^1$Institute  of Scientific Computing,  TU  Dresden,  01062  Dresden,  Germany,
  \\ 
  ${}^2$Dresden Center for Computational Materials Science (DCMS), TU  Dresden,  01062  Dresden,  Germany\\
   ${}^3$Department of Materials Science and Engineering, City University of Hong Kong, Hong Kong SAR, China\\
       ${}^4$Department of Mechanical Engineering, The University of Hong Kong, Pokfulam Road, Hong Kong SAR, China\\
  ${}^5$Hong Kong Institute for Advanced Study, City University of Hong Kong, Hong Kong SAR, China\\}
\end{center}
\vspace{10pt}
\setcounter{equation}{0}
\setcounter{figure}{0}
\setcounter{table}{0}
\setcounter{section}{0}
\setcounter{page}{1}
\makeatletter
\renewcommand{\thesection}{S\Roman{section}}
\renewcommand{\theequation}{S\arabic{equation}}
\renewcommand{\thefigure}{S\arabic{figure}}
\renewcommand{\bibnumfmt}[1]{[S#1]}
\renewcommand{\citenumfont}[1]{S#1}

\twocolumngrid

\section{Numerical Schemes}
\label{sec:AppA}

The model discussed in Sect.~III and Sect.~IV may be integrated by considering relatively simple approaches. Consider a semi-implicit integration scheme for handling the non-linear term $H'(\varphi)$ by a one-iteration Newton method,
$H'(\varphi^{n+1}_{i,j}) \approx H'(\varphi^{n}_{i,j})+H''(\varphi^{n}_{i,j})(\varphi^{n+1}_{i,j}-\varphi^{n}_{i,j})$, 
with $n$ and $(i,j)$ indexing the time  and  2D space discretisations. 
For finite difference implementations, the semi-implicit integration scheme for the full anisotropic PF model of Sect.~III, namely  Eq.~(21), is
{\small
\begin{widetext}
\begin{equation}
\begin{split}
\varphi^{n+1}_{i,j}
=&\bigg(\frac{1}{1+\varepsilon^{-2}M(\phi_{i,j}) \gamma(\phi_{i,j}) \Delta \tilde{t} H''(\varphi^{n}_{i,j})}\bigg)\bigg(\varphi^n_{i,j}+ \varepsilon^{-1}M(\phi_{i,j}) \Delta \tilde{t} \bigg(-\varepsilon^{-1}\gamma(\phi_{i,j}) H'(\varphi^{n}_{i,j})+\varepsilon^{-1}\gamma(\phi_{i,j}) H''(\varphi^{n}_{i,j})\varphi^{n}_{i,j}+\\
&+\varepsilon\mathcal{D}\bigg[\gamma(\phi_{i,j})\mathcal{G}(\varphi_{i,j}^n)\bigg]+\varepsilon\mathcal{D}\bigg[
\gamma'(\phi_{i,j})
\bigg(
\begin{array}{c}
-\mathcal{G}^y(\varphi^n_{i,j}) \\ 
\mathcal{G}^x(\varphi^n_{i,j})
\end{array}
\bigg)
\bigg] \bigg) 
+\Delta \tilde{t} |\mathcal{G}(\varphi^n_{i,j})|
(\bar{\tau}_{\rm self}+\tau_{\rm ext}+\psi)
\bigg),
\\
\end{split}
\label{eq:scheme_full}
\end{equation}
\end{widetext}
}
\noindent where $\mathcal{G}$ and $\mathcal{L}$ are the discretised gradient and Laplacian, respectively, $\mathcal{D}[\mathbf{A}_{i,j}]=\mathcal{G}^x[A_{i,j}^x]+\mathcal{G}^y[A_{i,j}^y]$ and $
\phi_{i,j}=\arctan[(\mathcal{G}^y_{i,j}(\varphi^n_{i,j}))/(\mathcal{G}^x_{i,j}(\varphi^n_{i,j})+\delta)]$, 
and where $\delta=10^{-6}$ a small regularisation parameter to avoid numerical divergences. 
Good accuracy is obtained by employing a five point stencil  for the first and second derivatives entering $\mathcal{G}$, $\mathcal{D}$ and $\mathcal{L}$. 
This approach may be exploited for testing on uniform grids. 
To properly resolve the interface thickness we should use  $\Delta x_1=\Delta x_2=\Delta s=\varepsilon/10$.

\begin{figure}[b]
   \includegraphics[width=0.75\columnwidth]{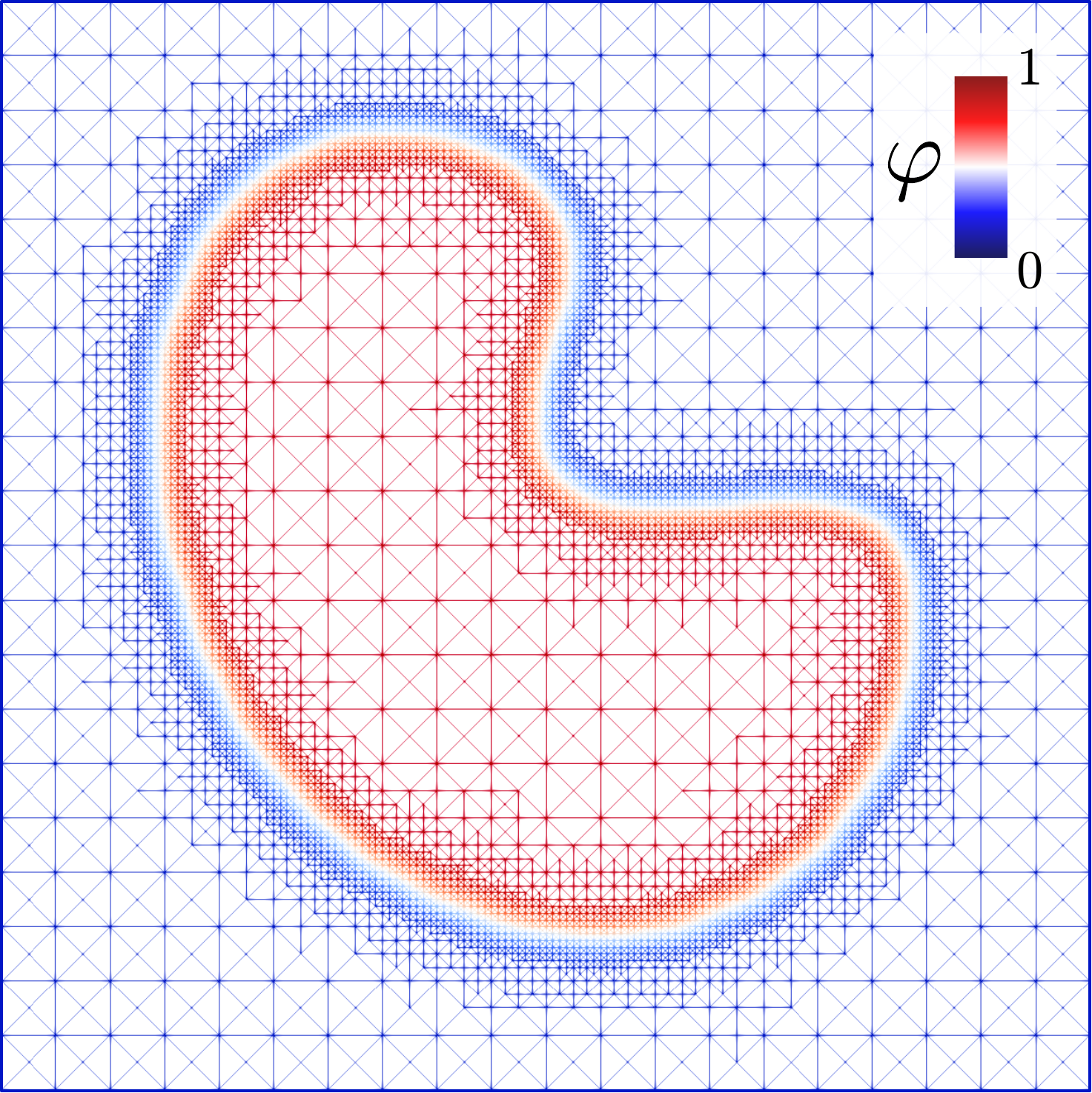} 
   \caption{Example of an adaptive mesh  employed in the finite element method implementation.}
    \label{fig:figureAPP}
\end{figure}

Rather, we employ spatial adaptivity, allowing  coarser resolution away from interfaces and  more efficient calculations in a Finite Element Method implementation with linear elements \cite{SMRatz2006}. 
An example refined mesh  is illustrated in Fig.~\ref{fig:figureAPP} (used for computations entering Fig.~2). 
In brief, a triangulation of the computational domain $\Omega$ at  timestep $n$, $\mathcal{T}_h^n$, is considered together with the finite element space of globally continuous, piecewise (linear) elements $\mathcal{S}_h^n=\{v_h \in X\ :\ v_h|_T \in P^1,\ \forall T \in \mathcal{T}_h^n \}$ with $X := \{ \xi \in \mathcal{H}^1(\Omega) : \xi_{\partial \Omega}\ \text{is periodic}\}$. Space discretisation is achieved by exploiting the weak form of Eq.~(21): i.e., find $\varphi_h^{n+1} \in \mathcal{T}_h^{n+1}$ such that
{\small 
\begin{widetext}
\begin{equation}
\begin{split}
\int_{\Omega}\frac{\varphi_h^{n+1}-\varphi_h^{n}}{\Delta t}\xi \ud \Omega =& - \int_{\Omega} \varepsilon^{-2}M(\hat{\mathbf{n}}^n)  \gamma(\hat{\mathbf{n}}^n)H''(\varphi_h^n) \varphi_h^{n+1} \xi \ud \Omega - \int_{\Omega} \varepsilon^{-2}M(\hat{\mathbf{n}}^n)  \gamma(\hat{\mathbf{n}}^n)\bigg( H'(\varphi_h^n)- H''(\varphi_h^n) \varphi_h^{n}\bigg) \xi \ud \Omega
\\
&- \int_\Omega M(\hat{\mathbf{n}}^n) \bigg( \gamma(\hat{\mathbf{n}}^n) \nabla \varphi_h^{n+1} + |\nabla \varphi_h^n|^2 \nabla_{\nabla \varphi_h} \gamma(\hat{\mathbf{n}}^n) \bigg)\cdot \nabla \xi \ud \Omega 
+ \int_\Omega|\nabla \varphi_h^n|(\bar{\tau}_{\rm self}+\tau_{\rm ext}+\psi)\xi \ud \Omega,
\label{eq:ac_fem_scheme}
\end{split}
\end{equation}
\end{widetext}
}
\noindent 
$\forall \xi \in \mathcal{S}_h^n$. This leads to a linear system of equations to solve for coefficients $\varphi_i^{n+1}$ of $\varphi_h^{n+1}=\sum_i\varphi_i^{n+1}\xi_i$. 
The numerical solution  is computed at every timestep exploiting the iterative, stabilized biconjugate gradient method (BiCGStab) and exploiting the adaptive finite element toolbox AMDiS\cite{SMVey2007,SMWitkowskiACM2015}. 
In the adaptive FEM,  we set $\Delta s$ to  the  element size in $\Sigma_\varepsilon$, with coarser elements elsewhere (see Fig.~\ref{fig:figureAPP}) and set $\Delta t = 10^{-1}\varepsilon$.
Simulations reported in this work can be obtained with both approaches briefly illustrated above. 
The FEM approach is most appropriate for extension to large length and time scales.

We compute $\bar{\tau}_{\rm self}(\mathbf{r})$ using $\tau_{\rm self}(s)$. 
The 0.5 isoline of $\varphi$ is extract via interpolation, resulting in a set of points $\mathbf{x}_i \in \Sigma$ discretising such a curve. $\tau_{\rm self}(s)\equiv \tau_{\rm self}(\mathbf{x}_i)$ is then computed on this curve as
{\small
\begin{equation}
\begin{split}
    \tau_{\rm self}(\mathbf{x}'_i)&=\sum_{\mathbf{x_i} \in \Sigma}  \bigg[\Delta S
\beta^{(1)} \frac{\ud x_2}{\ud s} \tau^{(1)}+\beta^{(2)} \frac{\ud x_1}{\ud s}
\tau^{(2)}\bigg]_{\mathbf{x}_i}, \\
\tau^{(m)} &= \frac{G}{2\pi(1-\nu)} 
\dfrac{x_m' - x_m}{\varrho_a^2} \bigg[
1 - \dfrac{2 \big(x_n' - x_n\big)^2}{\varrho_a^2}
\bigg],
\end{split}
\label{eq:s1}
\end{equation}}$(m,n)=(1,2)$ or $(2,1)$, $\ud x_j/\ud s=(x_{i+1}^j-x_{i-1}^j)(\Delta S)$, $\Delta S(\mathbf{x}_i) = \sqrt{|\mathbf{x}_{i+1} - {\mathbf{x}_i}| + |\mathbf{x}_{i}-{\mathbf{x}_{i-1}}|} $. 
Stable numerical integration of the differential problem (20) is achieved using an upwind scheme
\begin{equation}
\begin{split}
    (\bar{\tau}_{\rm self})^{n+1}_{i,j}=&(\bar{\tau}_{\rm self})^{n}_{i,j}+ \Delta p \bigg( 
    |n_x(\phi)| \mathcal{U}^x_\pm((\bar{\tau}_{\rm self})^{n}_{i,j}) \\
    &+ |n_y(\phi)| \mathcal{U}^y_\pm((\bar{\tau}_{\rm self})^{n}_{i,j}) \bigg),  \\
    \mathcal{U}^x_\pm(f_{i,j})=& \frac{f_{i+I,j}-f_{i,j}}{\Delta s}, ~ I=S(n_x)S(\varphi-0.5),\\
    \mathcal{U}^y_\pm(f_{i,j})=& \frac{f_{i,j+J}-f_{i,j}}{\Delta s}, ~ J=S(n_y)S(\varphi-0.5),\\
\end{split}
\label{eq:s2}
\end{equation}
where iterated  to convergence, i.e. $(\bar{\tau}_{\rm self})^{n+1}_{i,j} \approx (\bar{\tau}_{\rm self})^{n}_{i,j}$ everywhere. The fixed condition of $\bar{\tau}_{\rm self}(\mathbf{r})=\tau_{\rm self}(\mathbf{x})$ on $\Sigma$ is ensured by remapping nodes on $\Sigma$ into the 2D computational grid and keeping them fixed. This finite difference scheme is exploited also when considering non-uniform, adaptive grids (e.g. with adaptive FEM) similarly to [\onlinecite{SMSalvalaglioPRB2016}]. In all the simulations, we set $\Delta p =1$ and employ periodic boundary conditions. 
To properly account for periodic $\bar{\tau}(\mathbf{r})$, $\tau_{\rm self}$ should be computed by summing the elastic field of periodic image interfaces  (period equal to the computational domain size).

We employed these discretisation for the cases shown  in Sect.~III. 
The model reported in Sect~IV  can be considered by adapting the aforementioned scheme to obtain $N$ coupled systems to solve for $\varphi_k$ with $k=1,\cdots,N$. 
The additional term $\sqrt{2H({\varphi_k})}\lambda(\{\varphi_k\})$ is treated explicitly. 
The integrals \eqref{eq:s1} are computed for $\mathbf{x} \in \Sigma_{ij}$ and parameters $\beta_{ij}^{(m)}$. 
The integration scheme for the extension of $\tau(s)$, resulting in $\bar{\tau}(\mathbf{r})$, are\ performed by solving Eq.~\eqref{eq:s2} separately for each $\varphi_k$. 

\end{document}